\documentclass[traditabstract]{aa} % for the abstract without structuration
\usepackage{graphicx} \usepackage{epstopdf}
\usepackage{natbib} \usepackage{lscape}
\newcommand{\lsim}{\ \raise -2.truept\hbox{\rlap{\hbox{$\sim$}}\raise
    5.truept\hbox{$<$}\ }} \newcommand{\gsim}{\ \raise
  -2.truept\hbox{\rlap{\hbox{$\sim$}}\raise 5.truept\hbox{$>$}\ }}
\newcommand{\nodata}{...}  \newcommand{\sss}{VEGAS-SSS}
\newcommand{\gi}{$(g{-}i)$\ }
\begin{document}
   \title{VEGAS-SSS. A VST survey of Elliptical Galaxies in the Southern
     hemisphere: analysis of Small Stellar Systems}

   \subtitle{Testing the methodology on the globular cluster system
     in NGC\,3115}

   \author{Michele Cantiello\inst{1}\and Massimo Capaccioli\inst{2,3}\and
Nicola Napolitano\inst{3}\and
Aniello Grado\inst{3}\and
Luca Limatola\inst{3}\and
Maurizio Paolillo\inst{2,4} \and
Enrica Iodice\inst{3}\and
Aaron J. Romanowsky\inst{5,6}\and
Duncan A. Forbes\inst{7}\and
Gabriella Raimondo\inst{1}\and
Marilena Spavone\inst{3}\and
Francesco La Barbera\inst{3}\and
Thomas H. Puzia\inst{8,9}\and
Pietro Schipani\inst{3}
}
           
\institute{INAF Osservatorio Astr. di Teramo, via Maggini,-64100,
  Teramo, Italy \email{cantiello@oa-teramo.inaf.it} \and Dip. di
  Fisica, Universit\'a di Napoli Federico II, C.U. di Monte
  Sant'Angelo, Via Cintia, 80126 Naples, Italy \and INAF Osservatorio
  Astr. di Capodimonte Napoli, Salita Moiariello, 80131, Napoli, Italy
  \and Agenzia Spaziale Italiana - Science Data Center, Via del
  Politecnico snc, 00133 Roma, Italy \and University of California
  Observatories, 1156 High Street, Santa Cruz, CA 95064, USA \and
  Department of Physics and Astronomy, San Jos\'e State University,
  One Washington Square, San Jose, CA 95192, USA \and Centre for
  Astrophysics \& Supercomputing, Swinburne University, Hawthorn, VIC
  3122, Australia \and Institute of Astrophysics, Pontificia
  Universidad Cat\'olica de Chile, Avenida Vicu\~{n}a Mackenna 4860,
  Macul, 7820436, Santiago, Chile \and National Research Council
  Canada, Herzberg Institute of Astrophysics, 5071 West Saanich Road,
  Victoria, BC V9E 2E7, Canada }

%\institute{INAF Osservatorio Astronomico di Teramo, via M. Maggini
%  snc, I-64100, Teramo, Italy  \\ \email{cantiello@oa-teramo.inaf.it}
%           }

   \date{}

\authorrunning{Cantiello et al.} \titlerunning{VEGAS-SSS: NGC\,3115}

% \abstract{}{}{}{}{}
% 5 {} token are mandatory

  \abstract {We present a study of globular clusters (GCs) and other
    small stellar systems (SSSs) in the field of NGC\,3115, observed
    as part of the ongoing wide-field imaging survey VEGAS, carried
    out with the 2.6m VST telescope. We use deep $g$ and $i$
    observations of NGC\,3115, a well-studied lenticular galaxy with
    excellent scientific literature. This is fundamental for testing
    the methodologies, verifying the results, and probing the
    capabilities of the \sss. Leveraging the large field of view of
    VST allows us to carry out an accurate study of the distribution
    and properties of SSSs as a function of galactocentric distance,
    well beyond $\sim20$ galaxy effective radii, in a way not often
    possible. Our analysis of colours, magnitudes and sizes of SSS
    candidates confirms the results from the existing studies, some of
    which carried out with 8-10m class telescopes, and further extends
    them to previously unreached galactocentric distances, with
    comparable accuracy.  In particular, we find a colour bimodality
    for the GC population and a de Vaucouleurs $r^{1/4}$ profile for
    the surface density of GCs as for the galaxy light profile. The
    radial colour gradient of blue and red GCs already found, e.g., by
    the SLUGGS survey with Subaru and Keck data, is further extended
    out to the largest galactocentric radii inspected, $\sim65$ kpc.
    In addition, the surface density profiles of blue and red GCs
    taken separately are well approximated by a $r^{1/4}$ density
    profile, with the fraction of blue GCs being slightly larger at
    larger radii. We do not find hints of a trend for the red GC
    subpopulation and for the GC turnover magnitude to vary with
    radius, but we observe a $\sim0.2$ mag difference in the turnover
    magnitude of the blue and red GCs subpopulations. Finally,
    inspecting SSS sizes and colours we obtained a list of
    ultracompact dwarf galaxies and GC candidates suitable for future
    spectroscopic follow-up. In conclusion, the present study shows
    $i)$ the reliability of the methodologies developed to study SSSs
    in the field of bright early-type galaxies; and $ii)$ the great
    potential of the VEGAS survey to produce original results on SSSs
    science, mainly thanks to the wide-field imaging adopted.}
   \maketitle

%________________________________________________________________

\section{Introduction}

The study of the properties of old star clusters in and around
galaxies is one of the keystones for understanding the formation and
evolution of galaxies
\citep[][]{ashman92,forbes97,cote98,brodie06,tonini13}. Because of the
relative ease to detect them out to large galactocentric distances,
and of the lower complexity of their host stellar populations with
respect to massive galaxies, star clusters provide an accurate and
relatively straightforward tool to unveil the mechanisms that produced
the present distribution and evolutionary properties of stars in the
host galaxy.

The surroundings of massive galaxies are populated by a zoo of small
stellar systems (SSSs hereafter): globular clusters (GCs), extended
clusters (ECs), ultra compact dwarfs (UCDs), dwarf spheroidals
(dSphs), dwarf ellipticals (dEs), compact ellipticals (cE)
etc. \citep[see, e.g.,][and references therein]{forbes13}. The
characteristic magnitude, colours and half-light radii for some SSS
classes are given in Table \ref{tab_classes}. We emphasize that the
distinction between the different SSS types is sometimes not trivial,
and somewhat arbitrary, due to the lack of sharp distinction between 
the classes of SSSs as revealed, for example, by the scaling relations
of mass, radius, luminosity, central surface brightness, or velocity
dispersion
\citep[][]{drinkwater04,misgeld11,chiboucas11,mcconnachie12}. A
natural explanation to the lack of clear class-boundaries is that
there is not any. Indeed, the transformation processes occurring in
dense environments may cause the disruption or transformation of
massive SSSs, littering the galaxy field with the remains of disrupted
system: low mass SSSs, stellar streams,
etc. \citep[][]{bassino94,west95,dabrusco13,dabrusco14}.
 
Characterizing the properties of the wealth of SSSs in the potential
well of the host galaxy is fundamental for the understanding of their
origin, and is an important tool for gauging the growth of the galaxy
and, more in general, of cosmic structures.

In this context, the present study is dedicated to the analysis of
SSSs in NGC\,3115, and is the first of a series aimed at analyzing
SSSs in bright early-type galaxies, observed as part of the ongoing
imaging VST survey VEGAS (``VST survey of Elliptical Galaxies in the
Southern hemisphere'', distributed over many semesters; GTO-INAF
program, P.I. Massimo Capaccioli).

An overview of VEGAS, and of its scientific aims, is presented in
\citet{capaccioli14}. At completion, the survey will collect detailed
photometric information of $\sim100$ bright early-type galaxies, to
study the galaxy light distribution out to $\sim$15-20 effective
radii. These galaxy regions are still almost unexplored in the CCD
era, mainly because of the difficulties posed by the reduced detector
field-of-view. The coupling of a dedicated survey telescope, the VST
\citep{capaccioli11}, with a new generation wide-field optical imager,
the OmegaCAM \citep{kuijken11},
offers a great opportunity to investigate this issue. Similar studies
for the Northern hemisphere are being carried out for the Next
Generation Virgo Cluster Survey \citep[NGVS,][]{ferrarese12}, and the
MATLAS survey \citep{duc14}.

The specific aims of the \sss~ series is to study and characterize the
properties of the SSSs out to very large galactic radii, taking
  advantage of VEGAS imaging data. SSSs, especially the GC systems,
  have been studied for decades, and progress has been limited not so
  much by telescope collecting area but by field of view and by image
  quality (both to reduce contamination and to reduce the exposure
  times). Thus, the use of 8m and even 4m telescopes is not
  compelling, at least for the photometry. In this paper we show the
  original achievements possible with wide-field imaging from a 2.6m
  telescope.

So far, except for the already mentioned ongoing studies from the
  NGVS and MATLAS surveys, the SSSs field population of only a few
galaxies has been analyzed out to large galactocentric radii
\citep[][]{dirsch03,forbes11,usher12,blom12}, though typical studies
did not go much beyond $30\arcmin\times30\arcmin$, making difficult a
robust estimate of the total background contamination. Taking
advantage of the large field of view of the VST we will:
\begin{itemize}
\item analyze the photometry in $g$ and $i$ bands for candidates GCs,
  UCDs, ECs, dSphs, etc. Furthermore, at completion VEGAS will also
  include $r$ data for most of the targets, and $u$ for selected
  galaxies;
\item study the properties of various SSS populations as a function of
  galactocentric distance to limits presently unreached;
\item when possible, characterize the spatial extent of sources, with
  the specific purpose of increasing the efficiency in distinguishing
  between the various classes of SSSs;
\item provide catalogs of SSSs candidates essential for preparing
  spectroscopic follow-up campaigns based on samples suffering for low
  or, at least, controlled fore/background contamination. To this aim,
  \sss~ data covering the $u$ bands, possibly complemented with
  near-IR photometry, would be particularly efficient \citep{munoz14}.
\end{itemize}

Here, we present the analysis of the $g$ and $i$-band of the field
centered on \object{NGC\,3115}, with the aim of describing the data
reduction, the analysis tools and performances of the telescope, and
to anticipate the future exploitation of the survey. In
  particular, the present work will mostly focus on the properties of
  the GCs system in the galaxy. Throughout the paper we verify the
reliability of the methodologies used taking advantage of the large
amount of literature data available for NGC\,3115 (including results
from HST observations and 8-10m class telescopes), and present
original results on SSSs topics made possible by the use of the
large-format CCD mosaic. Indeed, the case of NGC\,3115, an isolated
lenticular galaxy, is particularly interesting for testing the
procedures used. Because of its proximity, the galaxy and its
satellites were targeted by many photometric and spectroscopic studies
\citep{hanes86,capaccioli87,kundu98,puzia00,puzia02,norris06,arnold11,usher12}. Moreover,
it is worth recalling that the GCs system of the galaxy is the first
one beyond the Local Group with confirmed bimodal metallicity
distribution, as shown by \citet{brodie12} from Calcium Triplet
analysis, and by \citet{cantiello14} using optical to near-IR
photometry \citep[see also][]{blake10a,yoon11a}.

The paper is organized as follows: the next section briefly describes
the observations and data reduction procedures. We introduce the data
analysis and the full catalog, providing the details of the
photometric and morphological study of SSSs candidates, in Section
\ref{sec_photo}. In Section \ref{sec_stat}, taking advantage of the
large field-of-view of the images, we study the properties of the GC
population versus galactocentric radius using a statistical background
decontamination method. Section \ref{sec_sizes} is dedicated to the
delicate issue of deriving SSS sizes. The final section provides a
summary of our main conclusions, and describes the perspectives for
the forthcoming \sss~ studies.

\begin{table}
\tiny
\caption{Typical properties of various classes of SSS}
\begin{tabular}{l c c c l}
\hline\hline
\hline
SSS Class  & $M_V$ (mag)        & $V{-}I$                 &   $R_H$ pc      & Reference         \\
%\multicolumn{2}{c}{Galaxy parameters}\\
\hline
GC         &$-11$ to $\gsim-5$     &         0.8-1.2        & 2-8            &  3, 6        \\ 
EC\tablefootmark{a}     &$-6 $ to $<-4$   &  $\sim$1.2        & 8-50           & 6, 7, 13      \\ 
UCD        &$-14$ to $-9$          &         0.7-1.3        &  8-100          & 1, 2, 4, 6      \\ 
dE         &$-16$ to $-12$         &         0.8-1.2        & 300-1000       &  5       \\ 
cE         &$-18$ to $-15$         &   $\sim1.2$            & 100-500        & 5, 11, 12    \\ 
dSph       &$-12$ to $\gsim-5$     &        0.8-1.2         & 50-1500        & 6, 8, 9, 10      \\ 
\end{tabular}
\label{tab_classes}
\tablebib{(1)~\citet{drinkwater04}; (2) \citet{mieske06c}; (3)
  \citet{harris96}; (4) \citet{mieske12}; (5)\citet{chiboucas11}; (6)
  \citet{brodie11}; (7) \citet{madrid11}; (8) \citet{mcconnachie12}; (9) \citet{karachentsev01}; (10) \citet{rejkuba06}; (11) \citet{misgeld09}; (12) \citet{mieske12}; (13) \citet{huxor05}.}
\tablefoottext{a}{Objects with similar luminosity and size have been also dubbed Faint Fuzzies \citep{larsen00b,peng06b,forbes13}.}
\end{table}

\section{Observations and data reduction}
\label{sec_obs}

The VST, VLT Survey Telescope, is a wide-field optical imaging
telescope with a 2.6-meter aperture, operating from the $u$ to the $z$
with a corrected field of view of 1 degree by 1 degree. Its single
focal plane instrument, OmegaCAM, is a large format (16k $\times$ 16k
pixels) CCD camera with a pixel scale of $0\farcs21~pixel^{-1}$.

The data reduction, including dither combination, vignetting and
exposure correction, astrometric solution and photometric calibration
was performed with the VST-Tube pipeline \citep{grado12}. Details
about the overall data quality can be found in
\citet{capaccioli14}. In particular the FWHM of the PSF varies for
$<0\farcs05$ across the field of view, and the internal astrometric
accuracy is $\sim0\farcs035$ (the $rms$ with respect to the USNO-B1
catalog is $\sim0\farcs2$).

In order to improve the analysis of the spatial extent of the sources
in the frame, we restricted our analysis to the imaging data with
average PSF FWHM$\leq0\farcs8$. With this choice the exposure time
is reduced by $\sim$30\% in $g$ and $\sim50\%$ in $i$ with respect to
the total integration time available.

\begin{figure*} 
   \centering
   \includegraphics[width=9cm]{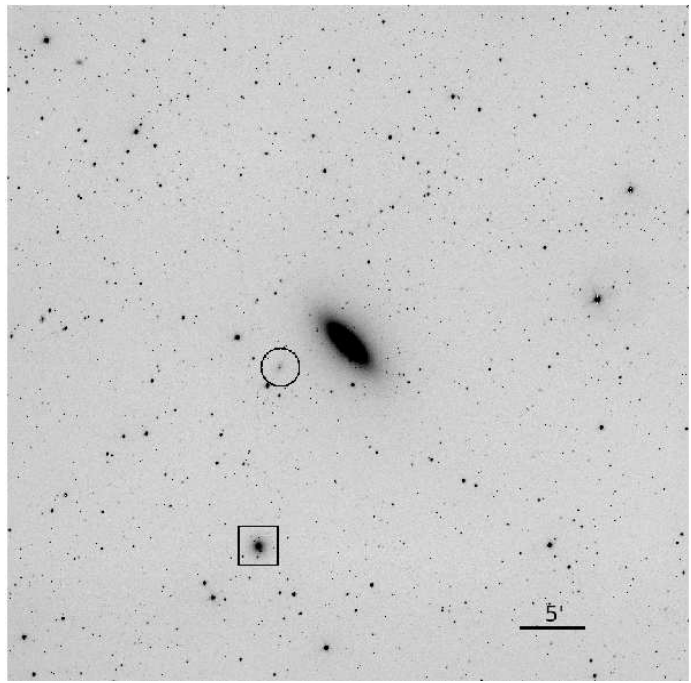} \caption{VST/OmegaCAM
     $g$-band image of NGC\,3115. North is up, East is left. The
     image size is $52.5\arcmin \times 52.5\arcmin$. The black square and
     circle mark the position of \object{NGC\,3115B} and
     \object{KK\,084}, respectively.}
   \label{ngc3115}
   \end{figure*}

Basic properties for the target and the optical observations are
listed in Table \ref{tab_props}. The $g$-band image of NGC\,3115 is
shown in Figure \ref{ngc3115}.

 %____________________________________ 

\begin{table}
\tiny
\caption{Main properties of NGC\,3115}
\begin{tabular}{l c}
\hline\hline
\hline
\multicolumn{2}{c}{Galaxy parameters}\\
\hline
RA(J2000)$^{\mathrm{1}}$                  &  10h05m14.0s \\
Dec(J2000)$^{\mathrm{1}}$                 &  -07d43m07s  \\
Galaxy Type$^{\mathrm{2}}$	               &   S0         \\      
Distance  adopted (Mpc)                & 9.4          \\
Absolute $B$-band magnitude$^{\mathrm{2}}$&  -19.9	      \\
$cz^{\mathrm{1}}$ (km/s, Heliocentric)    &  663$\pm$4   \\   
Mean $E(B{-}V)^{\mathrm{3}}$              &  0.042       \\    
Effective radius $R_{eff}$              &  57$\arcsec$  \\    
\hline
\multicolumn{2}{c}{Observations}\\
Filter (median FWHM)                & Exposure time (s)\\ 
$g$ ($\sim0.75\arcsec$)               &  2695            \\  
$i$ ($\sim0.72\arcsec$)               &  1250            \\ 
\hline \hline
\end{tabular}
\begin{list}{}{}
\item[$^{\mathrm{1}}$] Data retrieved from NED, nedwww.ipac.caltech.edu 
\item[$^{\mathrm{2}}$]  Hyperleda leda.univ-lyon1.fr
\item[$^{\mathrm{3}}$] \citet{sfd98} with \citet{sf11} recalibration
\end{list}
\label{tab_props}
\end{table}

Given our purpose of studying SSSs, we need to minimize the
contamination due to the presence of the light from NGC\,3115. To
model and subtract the galaxy, we used the ISOPHOTE/ELLIPSE task in
IRAF/STSDAS \citep{jedrzejewski87}\footnote{IRAF is distributed by the
  National Optical Astronomy Observatory, which is operated by the
  Association of Universities for Research in Astronomy (AURA) under
  cooperative agreement with the National Science Foundation.}. The
modeling failed to match the central thick disk region, which implied
poor detection of the sources within the inner $\sim 2\arcmin$
area\footnote{A test with GALFIT \citep{peng02}, a further program
  designed for modeling two-dimensional brightness profiles, also
  failed the modeling of the central galaxy regions. We also obtained
  a galaxy-subtracted frame as described in \citet{jordan07} and
  \citet[][which modeled NGC\,3115 from near-IR
    data]{cantiello14}. Such method uses the SExtractor spline
  background derived from the image logarithm, and provided very flat
  residuals. However, the latter procedure affects badly the shape of
  slightly extended objects, thus it is not suitable for the purposes
  of the present study.}. However, the central regions of the galaxy
have been accurately inspected using a $\sim10\arcmin \times 7\arcmin$
mosaic obtained with the ACS camera on board of the {\it Hubble Space
  Telescope} \citep[HST;][]{jennings14}. The ACS study relies on data
with similar wavelength coverage and $g$-band depth with respect to
the ones used here. Given the higher resolution of HST data, we do not
make any attempt to recover the sources in the central
$\sim3.5\arcmin\times1.5\arcmin$ poorly modeled regions along the
galaxy major axis.

\section{Photometry and size estimates}
\label{sec_photo}
To produce a complete catalog of all sources present in the VST field
of view, we run SExtractor \citep{bertin96} on the galaxy-model
subtracted frame, independently for each filter.

We obtained aperture magnitudes within a 6 pixel diameter aperture
($\sim1\farcs26$ at OmegaCAM resolution), and applied aperture
correction to infinite radius. The aperture correction, derived from
the analysis of the curve of growth of bright isolated point-like
sources \citep[see][for more details]{cantiello05,cantiello11b}, is
$0.52\pm0.01$ and $0.46\pm0.01$ mag in $g$ and $i$, respectively. For
extended sources, i.e. sources spatially more extended than the
instrumental FWHM of the PSF (see below), we used the SExtractor
Kron-like elliptical aperture magnitude.  Finally, the photometric
catalogs in the two bands were matched adopting $0\farcs5$ matching
radius. The final photometric catalog contained $\sim$47000 sources.

%_______DUST MAP____________________________ 
  \begin{figure*} 
\centering
\includegraphics[width=9.6cm]{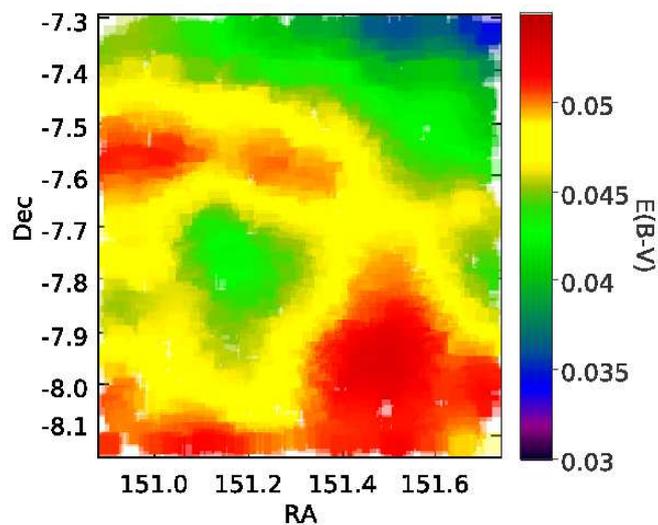}
  \caption{Extinction map over the area analyzed. The colorbar shows
    the extinction to color mapping.}
   \label{dust_map}
\end{figure*}

Due to the large areal coverage, there is a non-negligible variation
of Galactic extinction from one side to the other of the field
($\Delta A_g\sim0.07$ mag). We obtained the local extinctions from the
dust maps by \citet[][]{sfd98} and used the reddening factors from
\citet{sf11}. The final extinction map is shown in Figure
\ref{dust_map}. All further colours and magnitudes in the paper are
corrected for extinction unless otherwise stated. Other details on
the photometric properties of the images analyzed (completeness an
limiting magnitudes) are given in Appendix \ref{appendix_cc}.

\begin{figure*} 
\centering
\resizebox{1.0\textwidth}{!}{\includegraphics{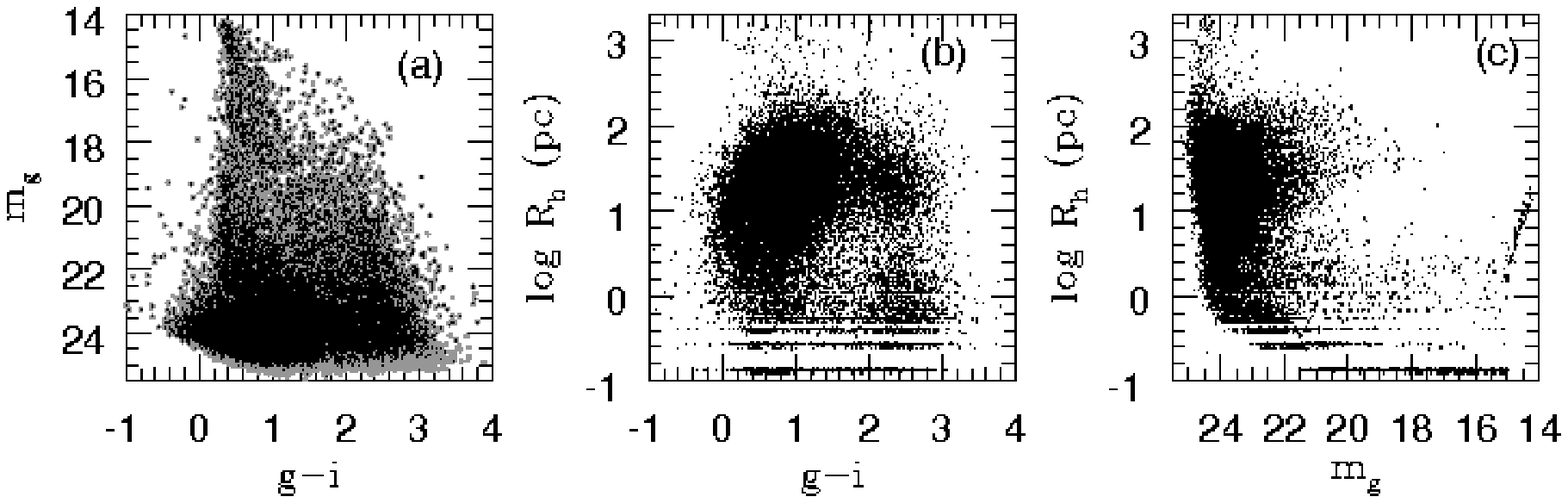}}
  \caption{Photometry and sizes of SSS candidates in NGC\,3115. Panel
    $(a)$: colour magnitude diagram for the full sample of $g$ and $i$
    matched sources (gray dots), and for the sources with size
    estimates (black points). Panel $(b)$: apparent size versus colour
    diagram. Panel $(c)$: apparent size versus magnitude diagram.}
   \label{colmag}
\end{figure*}

The colour magnitude diagram of the full sample of $g$ and $i$ matched
sources is shown in Figure \ref{colmag} (panel $(a)$).

%_______DUST MAP____________________________ 

It is very important to emphasize that the selection of SSSs based on
one single colour, the \gi, is inherently uncertain, resulting in a
catalog with large fractions of contaminating sources
\citep[foreground stars and background galaxies,][]{durrell14}. The
selection with a further optical color would certainly reduce the
fraction of contaminants, especially if $u$ band photometry is
available. However, a contaminant-free catalog based on optical
photometry is basically unattainable. It is useful to highlight,
though, that the coupling of optical data with just one near-IR band
is very effective in reducing the fraction of contamination to the GC
and UCD catalogs to less than $\sim5$\% \citep{munoz14}.

To partly overcome the problem of selecting SSSs relying on only one
optical colour, one can use statistical decontamination techniques
(see Section \ref{sec_stat}), and/or add a further selection
criterion: the physical extent of the source (Table
\ref{tab_classes}). The methodology that we will adopt to derive
objects sizes is described below, while the effectiveness and the
practical issues in using object sizes as a selection parameter will
be discussed in Section \ref{sec_sizes}.

\subsection{Size and shape measurements as compactness criterion}
\label{sec_measize}

As shown in Table \ref{tab_classes}, if one can estimate the
half-light radius $R_h$ of SSSs then the objects shape can be used
together with photometric properties to classify the system.  However,
size measurements can be very challenging, especially with
ground-based imaging data. Furthermore, in general one can only
measure angular sizes which, to be transformed in linear scale,
require the previous knowledge of the object distance. In spite of
this, angular sizes and shapes have been estimated for a large sample
of SSSs in different environments and with various ground- and
space-based telescopes
\citep[e.g.][]{larsen99,larsen03,jordan04b,cantiello07c,caso13,puzia14}. In
what follows we describe how object sizes have been estimated for
objects in the \sss~ fields.

Given the difficulty posed by the task, to estimate the intrinsic size
of a source exceeding some instrumental-dependent size limit, specific
tools have been designed and implemented to analyze the light profiles
of sources with intrinsic sizes comparable or slightly smaller than
the instrumental PSF.  For \sss~ we choose to adopt
\textsf{Ishape}\footnote{The software can be downloaded at
  http://baolab.astroduo.org/. For the present work we used the
  release 0.94.1d.} to obtain structural parameters (in particular
$R_h$ and  the axis ratio $b/a$) of SSSs. \textsf{Ishape} is
optimized for modeling the light distribution for marginally resolved
sources down to 1/10 of the FWHM of the PSF \citep{larsen99,larsen00}.
In such context, NGC\,3115 is one of the most attractive targets in
the survey, being also one of the nearest. At the adopted distance of
9.4 Mpc \citep[][using the updated calibration zeropoint from
  Cantiello et al., 2013]{tonry01}, and given the FWHM of the images
(Table \ref{tab_props}), \textsf{Ishape} can be used to determine the
physical extent of objects with $R_h\geq3.5$ pc. For the reasons
explained in Section \ref{sec_sizes}, we will take into account also
objects down to $R_h\sim2$ pc.  The measurement of source size below
the FWHM is particularly demanding in terms of signal-to-noise ratio,
SNR, for this reason we have checked that the $i$-band data currently
on hand did not provide us of an adequate SNR, hence in the following
we will use the $R_h$ estimates derived from the $g$-band only.

 \textsf{Ishape} reaches a convergence for $\sim30000$ of the input $\sim47000$
 sources in the photometric catalog. The size-colour and
 size-magnitude plots for the sample of sources for which we have
 structural parameters are shown in Figure \ref{colmag} (panels $(b)$
 and $(c)$, respectively).

More details on \textsf{Ishape} runs are given in Appendix \ref{app_ishape}.

\subsection{Final catalog}
\label{sec_catalogue}

\begin{table*}
\scriptsize
\caption{Photometry and size estimates for the matched $gi$ catalogs. The full table is available in the electronic version of the journal.}
\centering
\begin{tabular}{c| ccc | cccccc|cccc|c}
\hline\hline
           & \multicolumn{3}{|c|}{Position}                 &     \multicolumn{6}{|c|}{Photometry}                                                       &  \multicolumn{4}{|c|}{\textsf{Ishape} results} & \\
ID         &       RA       &        Dec     &   $R_{gal}$ & $m_g$               &    $CS_g$   &  $m_i$             & $CS_i$   &  \gi      & $E(B{-}V)$ &  SNR &  FWHM                           &  $R_h$  & $b/a$ &  Note\tablefootmark{a} \\
           &   (J2000)      &       (J2000)  &  ($\arcmin$)  &      (mag)        &             &  (mag)             &          &           &           &       & ($\arcsec$)                   &  (pc)   &            &       \\
 (1)       &   (2)          &       (3)      &  (4)          &      (5)          & (6)         &  (7)               & (8)      & (9)       & (10)      & (11)  & (12)      )                   &  (13)   & (14)       & (15)  \\
\hline
         4 & 151.040149 & -8.155928 &    30.7   &  25.35$\pm$   0.20 &    0.45 &     24.10$\pm$   0.23 &    0.56 &  1.250 &  0.053 &  \nodata   & \nodata &  \nodata   & \nodata    &    0 \\ 
         6 & 151.484945 & -8.155959 &    28.3   &  24.51$\pm$   0.09 &    0.70 &     23.91$\pm$   0.19 &    0.72 &  0.602 &  0.052 &  \nodata   & \nodata &  \nodata   & \nodata    &    0 \\ 
        92 & 151.348259 & -8.155089 &    26.3   &  23.97$\pm$   0.10 &    0.02 &     21.86$\pm$   0.08 &    0.06 &  2.109 &  0.051 &    18.6 &    0.95$^{ 0.33}_{-0.95}$ &    12.5 &  0.85 &    0 \\ 
        93 & 151.091210 & -8.154886 &    29.2   &  24.26$\pm$   0.12 &    0.39 &     23.05$\pm$   0.14 &    0.38 &  1.210 &  0.053 &    11.4 &    1.14$^{ 0.11}_{-1.14}$ &    15.0 &  0.86 &    0 \\ 
       293 & 151.571415 & -8.153376 &    30.4   &  21.78$\pm$   0.02 &    0.03 &     21.27$\pm$   0.04 &    0.03 &  0.502 &  0.051 &   106.7 &    2.49$^{ 0.03}_{-0.09}$ &    34.8 &  0.97 &    1 \\ 
       993 & 151.117979 & -8.143415 &    27.9   &  21.86$\pm$   0.02 &    0.03 &     21.19$\pm$   0.04 &    0.04 &  0.675 &  0.053 &    94.8 &    1.74$^{ 0.02}_{-0.10}$ &    22.0 &  0.78 &    1 \\ 
      1396 & 150.878854 & -8.138160 &    35.8   &  22.37$\pm$   0.02 &    0.86 &     21.67$\pm$   0.03 &    0.26 &  0.700 &  0.052 &    64.3 &    0.53$^{ 0.08}_{-0.01}$ &     6.7 &  0.77 &    2 \\ 
\hline\hline
\end{tabular}
\label{tab_catalog} 
\tablefoottext{a}{0: source common to both $g$ and $i$ cataloges without \textsf{Ishape} data, or rejected from the {\it reference} and {\it best} samples; 1: source in the {\it reference} sample; 2: source in the {\it best sample}.}
\end{table*}

The final catalog resulting from the colour and size/shape criteria is
given in Table \ref{tab_catalog}. The catalog contains the full list
of $\sim47000$ sources matched in the $g$ and $i$ catalogs. For each
source, the following parameters are reported: (1) \sss~ ID, (2) and
(3) right ascension and declination (J2000), (4) galactocentric
distance, (5) $g$ magnitude and error, (6) SExtractor CLASS\_STAR
parameter in the $g$ band $CS_g$, (7) $i$ magnitude and error, (8)
SExtractor CLASS\_STAR parameter in the $i$ band $CS_i$; (9) \gi
color; (10) local reddening; (11) signal-to-noise ratio from
\textsf{Ishape}; (12) FWHM of the source; (13) effective radius; (14)
object minor to major axis ratio ($b/a$); (15) notes.  The
absolute value of $R_h$ in pc depends on the distance adopted, thus it
is wrong for all unknown contaminating fore/background sources. In
Section \ref{sec_sizes} we will discuss the percentage of
contamination expected on the basis of comparison with data from the
literature.

\section{GC population properties as a function of galactocentric distance: statistical decontamination of the sample}
\label{sec_stat}

In this section, we analyze the colour and magnitude distribution of
SSSs in the field of NGC\,3115. In particular, because they dominate
the SSS population in the galaxy core, we focus on GCs, making use of
statistical decontamination of background sources. To have a better
statistics for the background subtraction, we use the entire \sss~
catalog of $g$ and $i$ matched sources ($\sim47000$ objects), and
select as good GCs candidates sources: $a)$ in the colour range for
$0.4 \leq(g{-}i)\leq1.25$ mag
\citep[e.g.][]{faifer11,kartha13,vanderbeke14}; $b)$ maximum
photometric error $\Delta (g{-}i)=0.15$ mag for colour analysis
($\Delta m_g=0.5$ mag for magnitude analysis); $c)$ SExtractor
star-galaxy $\langle CS\rangle\geq0.2$, to avoid contamination from
sources that are trivially background galaxies; $d)$ and $m_g\geq18$
mag, i.e. sources $\sim4~\sigma_{TOM}$ brighter than $m_g^{TOM}$ are
not taken into account (see below). Thus, for the analysis presented
in this section we do not apply any restriction on $R_h$.

\subsection{Background determination}
\label{sec_back}

Our approach relies on the assumption that all sources beyond some
limiting galactocentric radius, $R_{bg}$, are foreground or background
contaminants, with a uniform spatial distribution over the field. In
particular, we adopted $R_{bg}=23\arcmin$ (see also Section
\ref{sec_sizes}), corresponding to $\sim 65$ kpc at the distance of
the galaxy. Taking as reference the GCs systems in the \object{Milky
  Way} and \object{M\,31}, we estimate that a fraction of $\sim2-3\%$
GCs brighter than $m_g\sim25$ mag (the approximate 90\% limiting
magnitude in $g$, see Figure \ref{colmag}, and Appendix
\ref{appendix_cc}) might still be in the background sample because of
their large galactocentric distances $R_{gal}\gsim65$ kpc. More in
detail, the Galaxy has seven GCs, over 157, at $R_{gal}\gsim$65 kpc
\citep[][2010 release]{harris96}, only three of them are brighter than
the detection limit of our photometric catalog. This implies that, if
placed at the distance of NGC\,3115, and for random viewing angles,
$\lsim$2\% of the MW GCs would be included in the background
sample. The GCs catalog of M\,31 by \citet[][RBCv5]{galleti04},
selected using optical to near-IR colour cuts \citep{munoz14},
contains 447 GC candidates none of which at galactocentric distance
larger than 35 kpc. On the other hand, \citet{huxor14}, using the
CFHT/MegaCam data of the PAndAS survey, discovered 59 new GCs at large
galactocentric distances in Andromeda: 19 of them would be brighter
than our magnitude cut, and with a projected distance larger than 65
kpc from the galaxy center. This corresponds to a fraction $\sim3\%$
the total, assuming a total population of at least 700 GCs (638 from
RBCv5, 59 from PAndAS).

Adopting $R_{bg}=23\arcmin$ means that $\sim40\%$ of the image is used
for the analysis of contamination. The possible future addition of
further bands will allow to increase the inner radius for the
selection of GCs (more in general, of SSS satellites), allowing to use
a smaller fraction of the detector to characterize the contamination.

Under such assumption, the difference between the surface density at
galactocentric distance $R_{gal}\leq R_{bg}$ and $R_{gal}>R_{bg}$
gives the residual density of sources in NGC\,3115, mainly GCs.

We proceeded as follows. We first estimated the surface density of
background objects per square arcminute at given colour (or
magnitude), $\Sigma_{bg}$, adopting the selection criteria $(a)-(d)$
given above, plus the galactocentric distance. Then, the total surface
density of objects within elliptical concentric regions,
$\Sigma(R_{gal}\leq R_{bg})$, is estimated using the same criteria on
colour (or magnitude), adopting different inner radii, starting from
$R_{gal}=2\arcmin$ out to $R_{gal}=R_{bg}$, with $1\arcmin$ steps. The
geometry of the ellipses, with constant ellipticity $\epsilon=0.5$ and
position angle $PA=45^{\circ}$, is assumed according to the results of
\citet{capaccioli14} \citep[see also][]{arnold11,jennings14}. In the
following, $R_{gal}$ is the semi-major axis, if not stated otherwise.

The overdensity of sources at given colors (or magnitude) associated
with NGC\,3115 is finally estimated as the difference
$\Sigma_{Host}\equiv\Sigma(R_{gal}\leq R_{bg})-\Sigma_{bg}$.

\subsection{Colour Distribution}
\label{sec_colour}
 %______________________________________ 
  \begin{figure*} 
    \centering
  \includegraphics{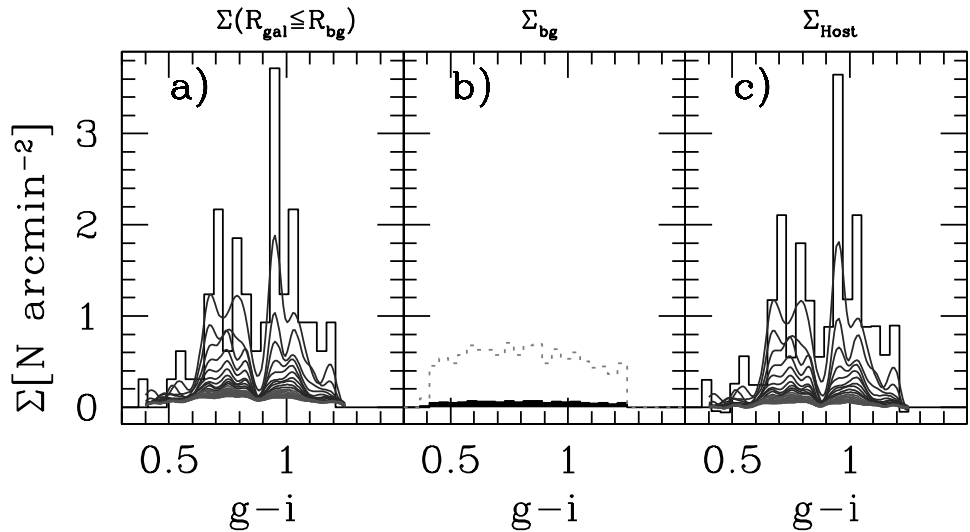}
  \caption{Surface density histograms versus color. Panel $(a)$:
    Surface density for sources within $R_{gal}\leq R_{bg}$. Darker
    colour refers to areas with smaller galactocentric radii. Panel
    $(b)$: same as left, but for background sources at
    $R_{gal}>R_{bg}$. For sake of clarity dotted histogram shows the
    density histogram times a factor 10. Panel $(c)$: residual surface
    density, $\Sigma(R_{gal}\leq R_{bg})-\Sigma_{bg}$.}
   \label{gcdens_gi}
   \end{figure*}

The panels in Figure \ref{gcdens_gi} show the density histograms
$\Sigma(R_{gal}\leq R_{bg})$, $\Sigma_{bg}$, and $\Sigma_{Host}$
versus colour (from left to right, respectively). In each panel,
darker colour refers to regions with smaller inner radii. The
histograms after the first innermost radius have been smoothed for
sake of clarity. In the first panel of the figure, the density
distribution shows the presence of a dip at \gi$\sim0.9$ mag, and two
well defined peaks at \gi$\sim0.75$ and $1.00$ mag whose prominence
decreases, but does not go to zero, as larger radii are considered.

For background sources (Figure \ref{gcdens_gi}, $\Sigma_{bg}$ middle
panel) the density distribution does not show relevant features, and
appears nearly flat over the colour interval shown. As expected, the
colour distribution of the difference diagram (Figure \ref{gcdens_gi},
$\Sigma_{Host}$ right panel) shows two distinct color peaks at all
radii.

To investigate the properties of the colour distributions in panel $(c)$
at each given radius, we used the ``Gaussian mixture modeling'' code
\citep[GMM,][]{muratov10}\footnote{GMM uses the likelihood-ratio test
  to compare the goodness of fit for double-Gaussians versus a
  single-Gaussian.  For the best-fit double model, it estimates the
  means and widths of the two components, their separation DD in terms
  of combined widths, and the kurtosis of the overall distribution. It
  also provides uncertainties based on bootstrap resampling.  In
  addition, the GMM analysis provides the positions, the relative
  widths, and the fraction of objects associated with each
  peak.}. More in details, we randomly populated the difference
distribution, $\Sigma_{Host}$, with a fixed number of sources
($N_{sim}\sim1500$), and then run the GMM code on the repopulated
sample.

\begin{figure}[h] 
  \resizebox{\hsize}{!}{\includegraphics{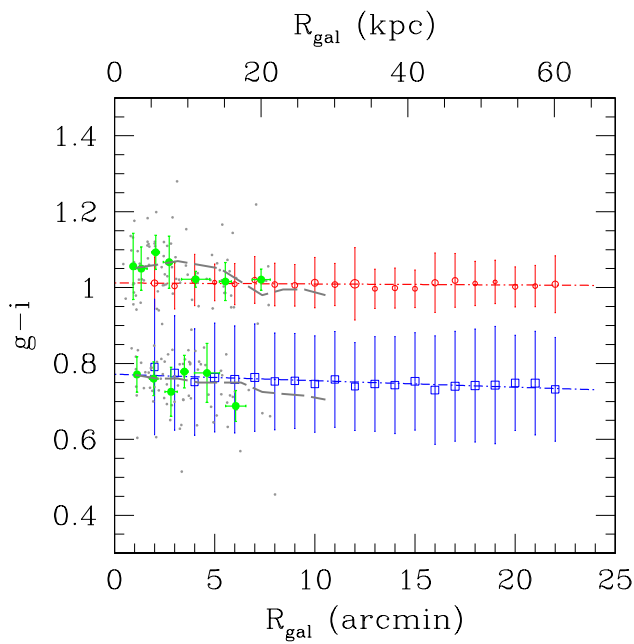}}
  \caption{Position and width of the blue and red GCs (blue squares
    and red empty circles, respectively) at different $R_{gal}$ as
    obtained from GMM. Symbol size is proportional to the fraction of
    objects associated with each peak. A fit to the data is shown with
    dot-dashed lines for both subpopulations. Gray dots show
    spectroscopic confirmed GCs from \citet{arnold11}. Full green
    points mark the running mean (median from equal number of data) of
    gray points. Gray long-dashed lines mark the rolling fits of the
    blue and red GC peaks as derived by \citet{arnold11} obtained from
    a combination of spectroscopic and photometric selected GCs.}
   \label{arnold}
   \end{figure}

The results of the GMM run are given in Table \ref{tab_gmm}, where for
each $R_{gal}$ it is reported the position of the peak and width of
the blue and red distributions, as well as the fraction of GCs
associated with each populations (in parentheses). The total fraction,
in some cases, does not equal one, because of the presence of a minor
very red peak. Figure \ref{arnold} shows the positions of the blue and
red peaks, the standard deviation of each distribution, and the
fraction of objects associated with each peak (given by symbol size).

\begin{table}
\scriptsize
\caption{Results of GMM at various galactocentric radii.}
\centering
\begin{tabular}{c c c c}
\hline\hline
$ R_{gal} (\arcmin) $  & $ (g{-}i)_0^{blue} $ &  $ (g{-}i)_0^{red} $  &  $ N_{GC}^{sel} $ \\  
\hline
       2.0 &       0.79 $\pm$        0.18 (      0.64) &        1.01 $\pm$        0.07 (      0.30) &       68 \\ 
        3.0 &       0.78 $\pm$        0.15 (      0.69) &        1.00 $\pm$        0.06 (      0.26) &      150 \\ 
        4.0 &       0.75 $\pm$        0.14 (      0.64) &        1.02 $\pm$        0.07 (      0.30) &      204 \\ 
        5.0 &       0.76 $\pm$        0.14 (      0.74) &        1.01 $\pm$        0.05 (      0.19) &      268 \\ 
        6.0 &       0.76 $\pm$        0.14 (      0.70) &        1.01 $\pm$        0.05 (      0.23) &      315 \\ 
        7.0 &       0.76 $\pm$        0.14 (      0.70) &        1.02 $\pm$        0.06 (      0.25) &      364 \\ 
        8.0 &       0.76 $\pm$        0.13 (      0.65) &        1.02 $\pm$        0.06 (      0.29) &      400 \\ 
        9.0 &       0.75 $\pm$        0.13 (      0.65) &        1.01 $\pm$        0.06 (      0.29) &      454 \\ 
       10.0 &       0.75 $\pm$        0.13 (      0.65) &        1.00 $\pm$        0.05 (      0.23) &      518 \\ 
       11.0 &       0.76 $\pm$        0.13 (      0.66) &        1.00 $\pm$        0.06 (      0.24) &      562 \\ 
       12.0 &       0.75 $\pm$        0.13 (      0.67) &        1.00 $\pm$        0.06 (      0.27) &      612 \\ 
       13.0 &       0.75 $\pm$        0.13 (      0.69) &        1.00 $\pm$        0.05 (      0.22) &      661 \\ 
       14.0 &       0.75 $\pm$        0.14 (      0.69) &        1.00 $\pm$        0.05 (      0.21) &      718 \\ 
       15.0 &       0.75 $\pm$        0.13 (      0.70) &        1.00 $\pm$        0.05 (      0.23) &      768 \\ 
       16.0 &       0.74 $\pm$        0.14 (      0.69) &        1.01 $\pm$        0.08 (      0.29) &      833 \\ 
       17.0 &       0.75 $\pm$        0.14 (      0.70) &        1.02 $\pm$        0.07 (      0.28) &      890 \\ 
       18.0 &       0.75 $\pm$        0.13 (      0.69) &        1.00 $\pm$        0.05 (      0.23) &      963 \\ 
       19.0 &       0.75 $\pm$        0.13 (      0.69) &        1.00 $\pm$        0.05 (      0.22) &     1024 \\ 
       20.0 &       0.75 $\pm$        0.13 (      0.69) &        1.01 $\pm$        0.06 (      0.24) &     1086 \\ 
       21.0 &       0.75 $\pm$        0.14 (      0.70) &        1.00 $\pm$        0.05 (      0.22) &     1159 \\ 
       22.0 &       0.73 $\pm$        0.14 (      0.66) &        1.01 $\pm$        0.08 (      0.31) &     1247 \\ 
 \hline\hline
\end{tabular}
\label{tab_gmm}
\end{table}

In the table we also report the number of GC candidates selected
according to the $(a)-(d)$ selection criteria given above
($N_{GC}^{sel}$, column). Using the average surface density of
contaminants $\Sigma_{bg}=0.056\pm0.014~[N/arcmin^{2}]$, and the
$N_{GC}^{sel}$ listed, one can easily derive the expected number of
GCs corrected for contamination at each elliptical radius. As an
example, the area with $R_{gal}\sim6-8\arcmin$ roughly corresponds to
the ACS area inspected by \citet{jennings14}, and is expected to
contain $\sim310-390$ GCs, to be compared with the 360 candidates
found with ACS.

We note that the position of the two peaks and their width are
consistent at all radii inspected, and agree very well with the recent
similar analysis on the same host galaxy
\citep[][]{faifer11,usher12}. A closer inspection to the data in
Figure \ref{arnold}, and Table \ref{tab_gmm} reveals the presence of
important features. First, a colour-$R_{gal}$ correlation is observed
for the blue GC component (Pearson correlation coefficients
$r_{xy}\sim-0.8$), with a $\sim0.06$ mag colour difference between the
inner and outer region. There is no, or a very weak, colour-radius
correlation for the red GCs ($r_{xy}\sim-0.25$). Furthermore, the
fraction of sources in the red sub-population shows a slight but
significant decrease with respect to the blue one at large radii. The
width of both sequences is relatively stable with radius, with the
blue distribution being broader at all radii.

These properties support a scenario where blue GCs are associated with
the galaxy halo, while red ones are more centrally concentrated and
associated with the bulge stellar component in the galaxy
  \citep[][]{kp97,cote98,forte05,liu11}.

In order to study the population of GCs associated with NGC\,3115
excluding the GC contaminants from the neighboring fainter galaxies,
we also carried out several tests rejecting all GC candidates within
2-3$\arcmin$ from KK\,084 and NGC\,3115B. The first galaxy, KK\,084,
is a dSph with center at $R_{gal}\sim5.5\arcmin$ from NGC\,3115, and a
non-negligible population of GC candidates, having a specific
frequency $S_N\equiv N_{GC}~10^{0.4(M_V+15)}=10$
\citep[][]{harris81,puzia08}.  In spite of the relatively large $S_N$,
the net effect on the properties of the GC system in NGC\,3115 is
negligible.  None of the sources in NGC\,3115B falls in the elliptical
shaped area of NGC\,3115 inspected here.

 %______________________________________ 

In Figure \ref{arnold} we added the data from \citet{arnold11}, which
are part of the SLUGGS survey \citep{brodie14}. Gray circles in the
figure mark spectroscopic confirmed GCs, green dots mark the running
mean for gray dots. We find very good matching between the mean \sss~
colour obtained with the statistical decontamination approach
presented in this section, and the colour of the spectroscopically
confirmed GCs.

\citet{arnold11} also derived the radial profiles out to
$R_{gal}\sim10\arcmin$ combining the spectroscopic sample with a
photometric sample, corrected for contamination using {\it ``an
  iterative Monte Carlo scheme''} (gray lines in
Fig. \ref{arnold}). The matching of the \sss~ and SLUGGS colour
profiles for blue GCs is good at all common radii.

We note that at $R_{gal}\geq6\arcmin$ the colour profiles from
\citeauthor{arnold11} depend mostly on the properties of the
photometric sample, thus the transition appearing in both the blue and
red GCs profiles at $6\geq R_{gal}\geq8\arcmin$ is strongly weighted
toward the photometric sample.

For the blue GCs, the difference between the mean from VST and
\citeauthor{arnold11}'s colour of spectroscopically confirmed GCs at
$R_{gal}\leq6\arcmin$ is $\Delta\gi^{blue}<0.01$ mag. For the red GCs
component the difference is $\Delta\gi^{red}\sim0.03$ mag.

By coupling the spectroscopic and photometric samples (gray dashed
line), \citeauthor{arnold11} found that the red GCs are on average
bluer at larger galactocentric distances as for blue GCs. The presence
of a radial trend in the red GCs from SLUGGS data appears mostly
beyond $R_{gal}\sim6\arcmin$, where the photometric sample dominates
over the spectroscopic one. Furthermore, the red GCs profile is nearly
flat for $R_{gal}\leq5.5\arcmin$ and $R_{gal}\geq7.5\arcmin$, with a
$\sim0.07$ mag colour transition in between.

Overall, the radial colour profiles of GCs from the \sss~ and SLUGGS
are consistent if one takes into account the error envelopes, the
intrinsic width of the distribution at fixed $R_{gal}$ and the
different analysis approaches adopted.

The good matching appears even more strikingly if one takes into
account that the data from \citet{arnold11}, are obtained by coupling
$gri$ band imaging data from Suprime-Cam at the 8.2m Subaru telescope,
and spectroscopy from the 10m Keck-II telescope with DEIMOS.

In conclusion, the comparison shown in Figure \ref{arnold} provides a
strong evidence in support of the efficiency of the approach adopted
here to analyze the properties of the GC system out to more than
$\sim20$ galaxy effective radii. It also shows that original
  results are obtained, even with the photometry in only two
  passbands, when using the wide-field imaging data from the 2.6m VST
  telescope.

\subsection{Surface density profiles}
\label{sec_sbp}

\begin{figure}[h] 
  \resizebox{\hsize}{!}{\includegraphics{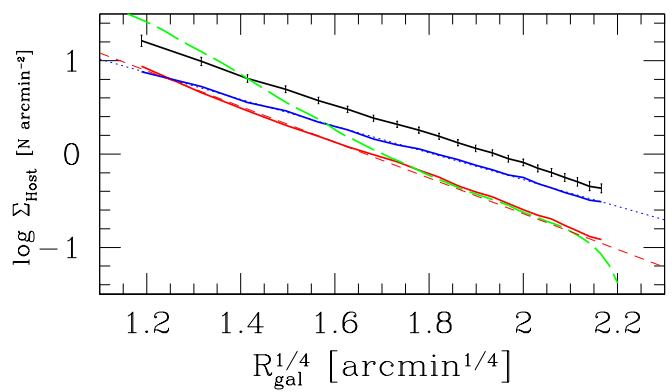}}
  \caption{Surface density profiles of blue, red, and total GC
    population (blue, red and black lines, respectively). The galaxy
    surface brightness profile in $g$ band from \citet{capaccioli14}
    is also reported with green long-dashed line. The linear fit to
    the surface density is shown with dotted lines (blue-dotted,
    red-dashed for the blue/red GCs, respectively). The scale of the
    galaxy profile is arbitrary.}
   \label{br_dens}
   \end{figure}

The radial profiles of the projected surface density for GC candidates
are shown in Figure \ref{br_dens}. The surface density at each radius
is obtained as the difference between the total density of sources
with $R_{gal}\leq R_{bg}$, and the background density. 

Taking advantage of the results obtained with GMM on the blue/red GCs,
we also analyzed the radial density profiles of the blue/red
subpopulations. Dividing the GCs into subpopulations, adopting a sharp
blue/red separation at $g{-}i=0.9$, the radial profile for the red GCs
appears steeper than that for the blue GCs. Moreover, both density
profiles follow very closely a $r^{1/4}$ de Vaucouleurs profile
(dotted lines), and both are shallower than the galaxy light profile,
showing a behavior similar to other galaxies \citep[e.g. NGC\,4636 and
  NGC\,3923,][]{dirsch05,norris12}.

 %______________________________________ 

 %______________________________________ 
  \begin{figure*} 
     \centering
\resizebox{1.0\textwidth}{!}{\includegraphics{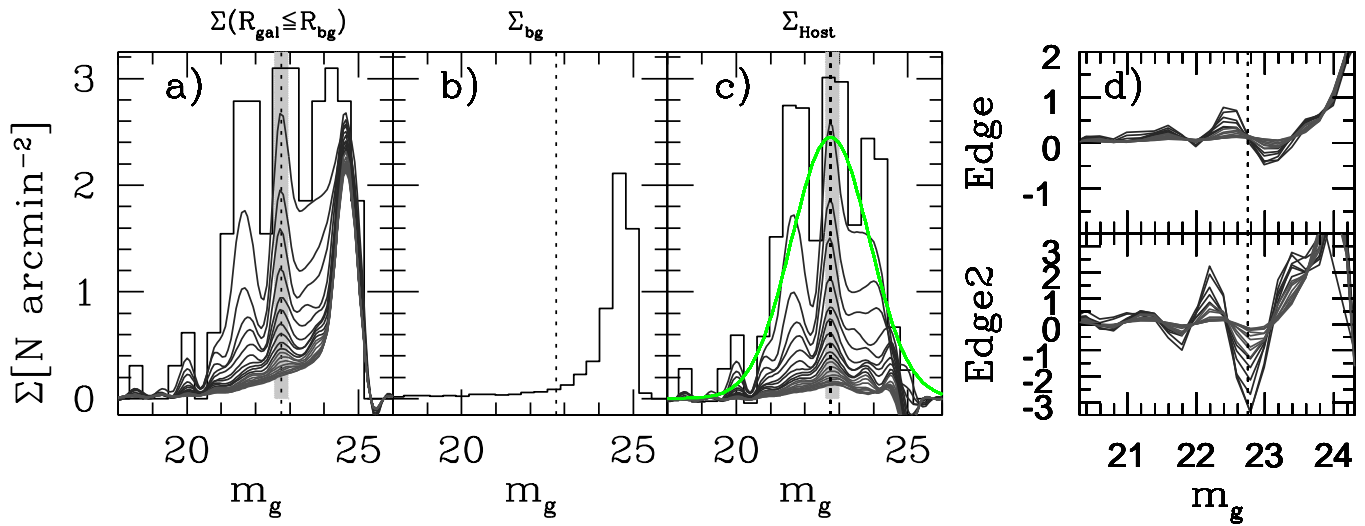}}
  \caption{Surface density histograms versus magnitude. Panels $(a)$
    to $(c)$: as in Figure \ref{gcdens_gi} but magnitude is used
    instead of colour. The vertical dotted line shows the position of
    the TOM. The $\sim0.2$ mag tolerance for the peak position (gray
    shaded area) is also shown as well as the best fit Gaussian to the
    GCLF (green line; $m_{g}^{TOM}=22.75$ mag,
    $\sigma_{TOM}=1.14$). Panel $(d)$: edge and second-run edge
    (Edge2) functions.}
   \label{gcdens_g}
   \end{figure*}

The steeper starlight gradient, compared with GCs density (blue or
total GC density), suggests that the GC system of NGC\,3115 extends
farther than the surface brightness profile of the galaxy halo. This
result is consistent with the general picture of the GC system being
spatially more extended than the host galaxy
\citep{harris91,harris00,forbes06,alamo12,kartha13}.

A further feature in Figure \ref{br_dens} is the matching of the
density profile for red GCs with the galaxy light profile at
$R_{gal}\geq 7.5\arcmin$ ($R^{1/4}_{gal}\geq1.65$), while the surface
density of GCs at smaller radii is slightly lower. Such depletion has
already been observed in galaxies brighter than NGC\,3115
\citep[e.g.][]{dirsch05,goudfrooij07}, and associated with higher
efficiency of GC-disruption mechanisms in the inner galaxy regions
\citep[dynamical friction, two-body relaxation and GC tidal
  shocking,][]{vesperini01,goudfrooij07}, suggesting that the galaxy
has undergone a relatively quiescent evolution, without major
star-forming events, which would have increased the inner density of
red GCs.

\subsection{Luminosity Function: GCLF}
Adopting the same approach used for colors in Section
\ref{sec_colour}, we analyzed the luminosity function of sources in
the field, with the specific purpose of inspecting the GC luminosity
function (GCLF) to independently estimate the galaxy distance modulus
\citep{harris01}, and further derive the position of the turnover
magnitude $m_g^{TOM}$ as a function of galactocentric distance.

Figure \ref{gcdens_g} shows the surface density distribution obtained
as described in previous section, with the difference that in this
case we used the total $g$ magnitude, instead of \gi colour. The GCLFs
derived are corrected for radial-dependent completeness as described
in Appendix \ref{appendix_cc}. Panel $(a)$ in the figure shows the
presence of various local maxima in the density distribution, whereas
the distribution of background sources, in panel $(b)$, has a power
law increase with a drop between $m_g\sim24$ and $25.5$ mag, due to
the completeness limit given by the adopted selection criteria. The
density distribution of sources in the host galaxy, shown in panel
$(c)$, reveals the presence of a major peak at $m_g\sim22.75$ mag.

To inspect the presence of a discontinuity in the luminosity function
due to the TOM, we adopt a quantitative method introduced by
\citet{lee93} to identify the position of the RGB Tip in Galactic
resolved GCs.  The results of such edge-detection method (based on the
Sobel filter, see Appendix \ref{appendix_cc}) are shown in panel $(d)$
of Figure \ref{gcdens_g}. Although the uncertainties in the surface
density, and their propagation in the definition in the edge and
second-edge functions are certainly large, the diagrams highlight the
presence of an inflection point (edge) and a maximum (edge2) around
$m_g\sim22.75$ mag as expected at the TOM (see Figure
\ref{simuledge}\footnote{Inspecting the edge functions two other
  possible TOM-point candidates are located at $m_g\sim22.2$ and 23.2
  mag (see Appendix \ref{appendix_cc}). However, both magnitude values
  are ruled out as TOM peak by the shape of the GCLF.}).

In Figure \ref{gcdens_g}, we also show the gaussian GCLF with
arbitrary peak normalization, assuming turnover TOM magnitude
$m_g^{TOM}\sim22.75$, with $\sigma_{TOM}=1.14$ derived from
\citet[][eq. 18]{jordan09}. A $\sim0.2$ mag tolerance area around
$m_g^{TOM}$ is also shown. Adopting the absolute value for the
turnover magnitude from the ACS Virgo Cluster Survey for galaxies with
$M_B<-18$, $M_g^{TOM}=-7.2\pm0.2$ mag, we estimate a distance modulus
$\mu_0=29.95\pm0.3$, in good agreement with the literature distance of
the galaxy (Table \ref{tab_props}).

Furthermore, thanks again to the large area inspected, we also probe
the variation of $m_g^{TOM}$ to large projected galactocentric
radii. The data in Figure \ref{gcdens_g} (panel $c$) reveal a TOM
essentially constant over the spatial scales inspected, as also found
in other galaxies \citep{jordan07}.

On the basis of the results shown in Figure \ref{gcdens_g} and in
Figure \ref{gcdens_gi}, we deduce a low contamination rate in the
regions within $R_{gal}\leq 8\arcmin$ (first six darker curves in the
figures), as the luminosity and colour surface density of
contaminants, $\Sigma_{bg}$ shown in the central panels, can be one
order of magnitude smaller than density in the inner regions. This
implies that the rate of contamination of the \sss~ catalog for
NGC\,3115 is quite low for the innermost $\sim8\arcmin$. As an
example, the background density at $m_g\sim 23$ mag is $\Sigma_{bg}
\sim0.3 [N/arcmin^{2}]$, while the density of sources within
$R_{gal}\leq 8\arcmin$ is $\sim3.5$ times larger, and
gets $\sim7$ times larger at $R_{gal}\leq 4\arcmin$. 

 %______________________________________ 
  \begin{figure*} 
     \centering
  \includegraphics[width=14cm]{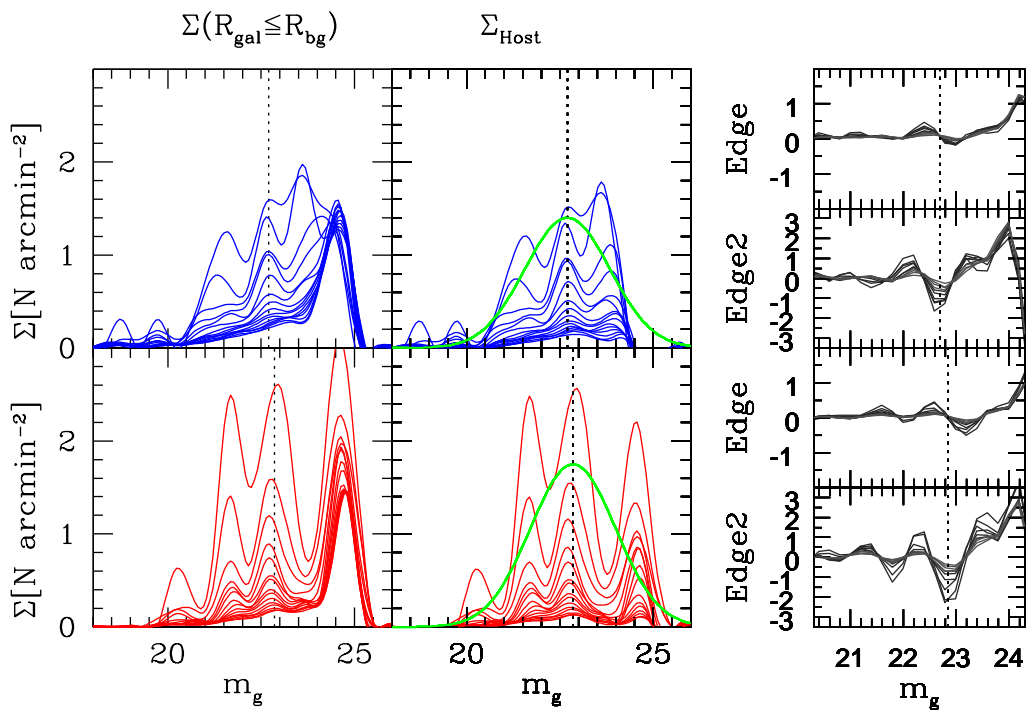}
  \caption{GCLF for the blue and red GC components analyzed
    separately. The upper panels, from left to right, show the total
    density distribution $\Sigma(R_{gal}\leq R_{bg})$, the density
    after subtracting for background contamination $\Sigma_{Host}$,
    and the Edge and Edge2 functions. The vertical dashed line marks
    the approximate position of the TOM, as obtained from the Edge
    functions. The green line shows a Gaussian with peak at
    $m_g^{TOM}=22.7$ mag. Lower panels: as upper ones, though for the
    red GC components, with $m_g^{TOM}=22.85$ mag.}
   \label{gcdens_br}
   \end{figure*}

We further inspected how the TOM differs between red and blue GCs, a
test that is not often possible and here feasible thanks to the large
area inspected. After dividing the blue/red GCs adopting a sharp
colour separation at \gi=0.9, we carried out the analysis described
above on the luminosity functions blue and red GCs. The results are
shown in Figure \ref{gcdens_br}, where the total luminosity function
$\Sigma(R_{gal}\leq R_{bg})$, the background corrected ones
$\Sigma_{Host}$, and the Edge/Edge2 diagrams are shown for the blue
and red GCs (upper and lower panels, respectively). Despite the
samples adopted are numerically smaller than before, the corrected
GCLF still shows the presence of a peak around the same $m_g^{TOM}$ of
the total GC population. By estimating the position of the TOM with
the Edge functions (right panels in the figure), the interesting point
here is that there appear to be a $\sim0.2$ mag offset between the TOM
of red and blue GCs, with the red system being fainter. From the point
of view of stellar population models, if the GCs mass function is
universal across metallicity, the Gaussian mean of the blue GCs is
expected to be brighter than that of the red one
\citep{ashman95,dicriscienzo06,jordan07}.  Observationally, our result
confirms previous evidences obtained from data with much smaller
surface coverage \citep{whitmore95b,puzia99,peng09}. Further
improvements on this will be allowed by the analysis of new galaxies
in the \sss~ sample, with the possible inclusion of $u$ and $r$ band
data in the SSSs selection process.

As a final comment, we highlight that the depth, in terms of absolute
magnitude, and the image quality for the other galaxies in the VEGAS
sample will be similar to the one inspected here, thus we expect that
using the tools presented here\footnote{Apart from the size/shape
  inspection, that will hardly be possible for GCs in distant
  galaxies, but doable for the UCDs.}, we will reliably analyze the
colour distributions and study GCs luminosity function for all other
targeted galaxies out to unreached galaxy effective radii. Moreover,
for objects at larger distances the background decontamination methods
described in this section will likely be more effective because of the
larger galactocentric distances inspected.

%____PHOTOMETRY COMPARISON ACS VST__________________________________________ 
   \begin{figure*}[ht]
   \centering
   \includegraphics[width=9cm]{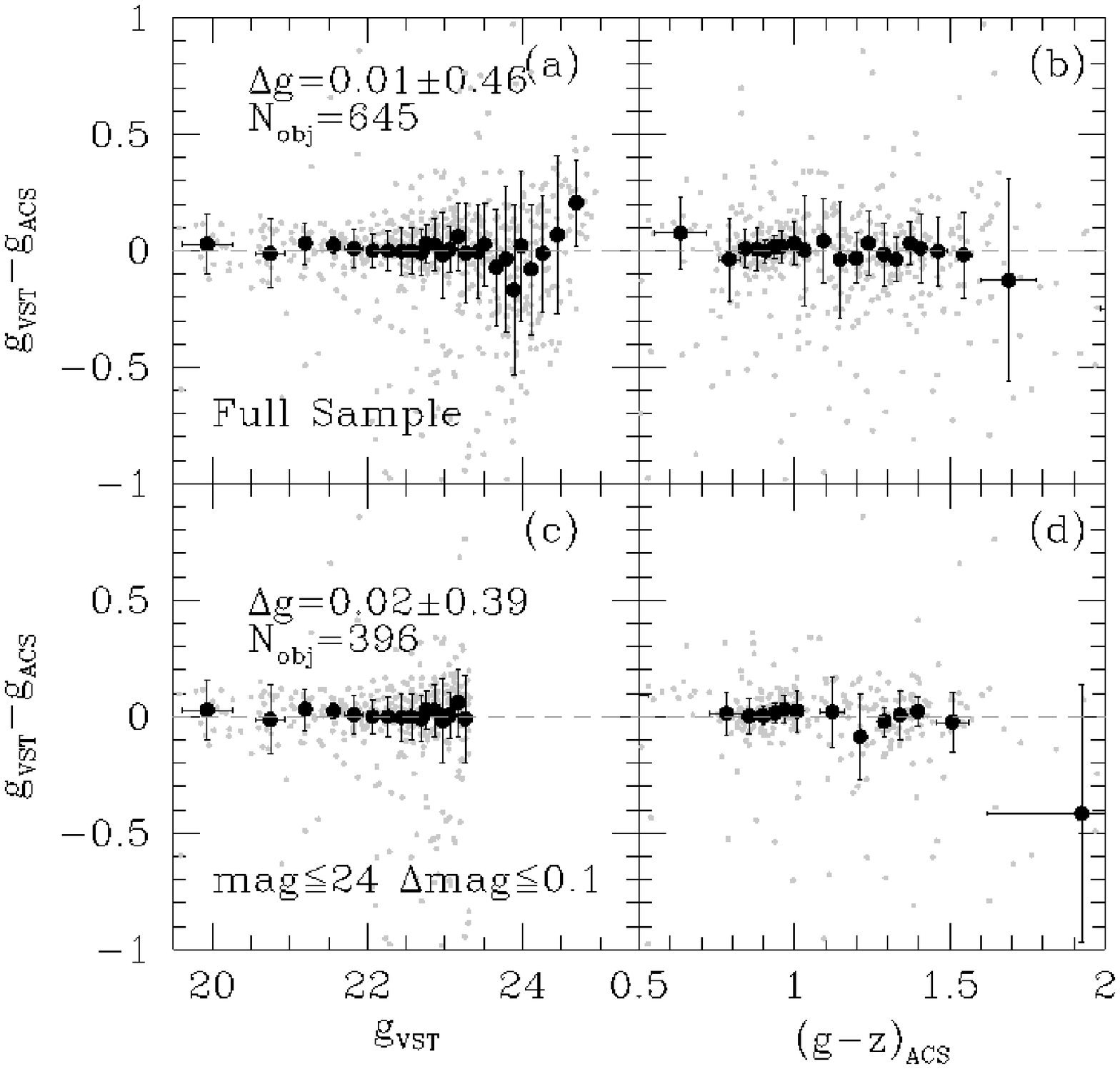}
   \includegraphics[width=9cm]{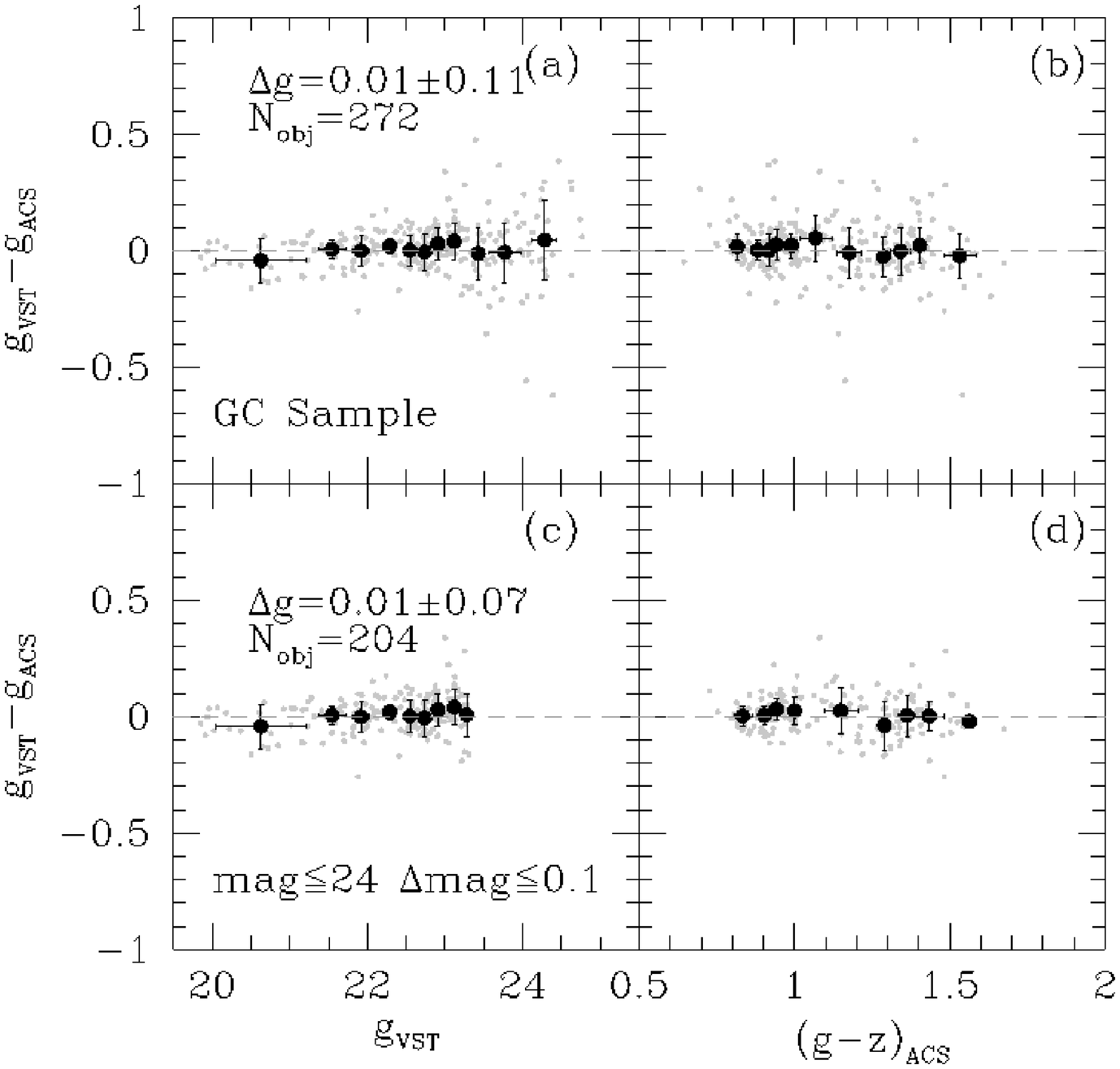}
   \caption{Magnitude comparison between \sss~ and ACS.  Left panels:
     $(a)$ and $(b)$ comparison for the full list of ACS to \sss~
     matching sources (gray dots). The running mean, and the
     corresponding $rms$, are shown with black circles and error
     bars. The median difference, the $rms$ and the number of objects
     matched are also labeled. Panels $(c)$ and $(d)$ same as upper
     panels, but for sources with magnitude and photometric error cuts
     as labeled. Right: as left panels, but only GC candidates in the
     ACS catalog are considered.}
   \label{acsvstmag}
   \end{figure*}
%

%____ISHAPE TESTS ACS COMPARISON__________________________________________ 
   \begin{figure*}
   \centering
  \includegraphics[width=9cm]{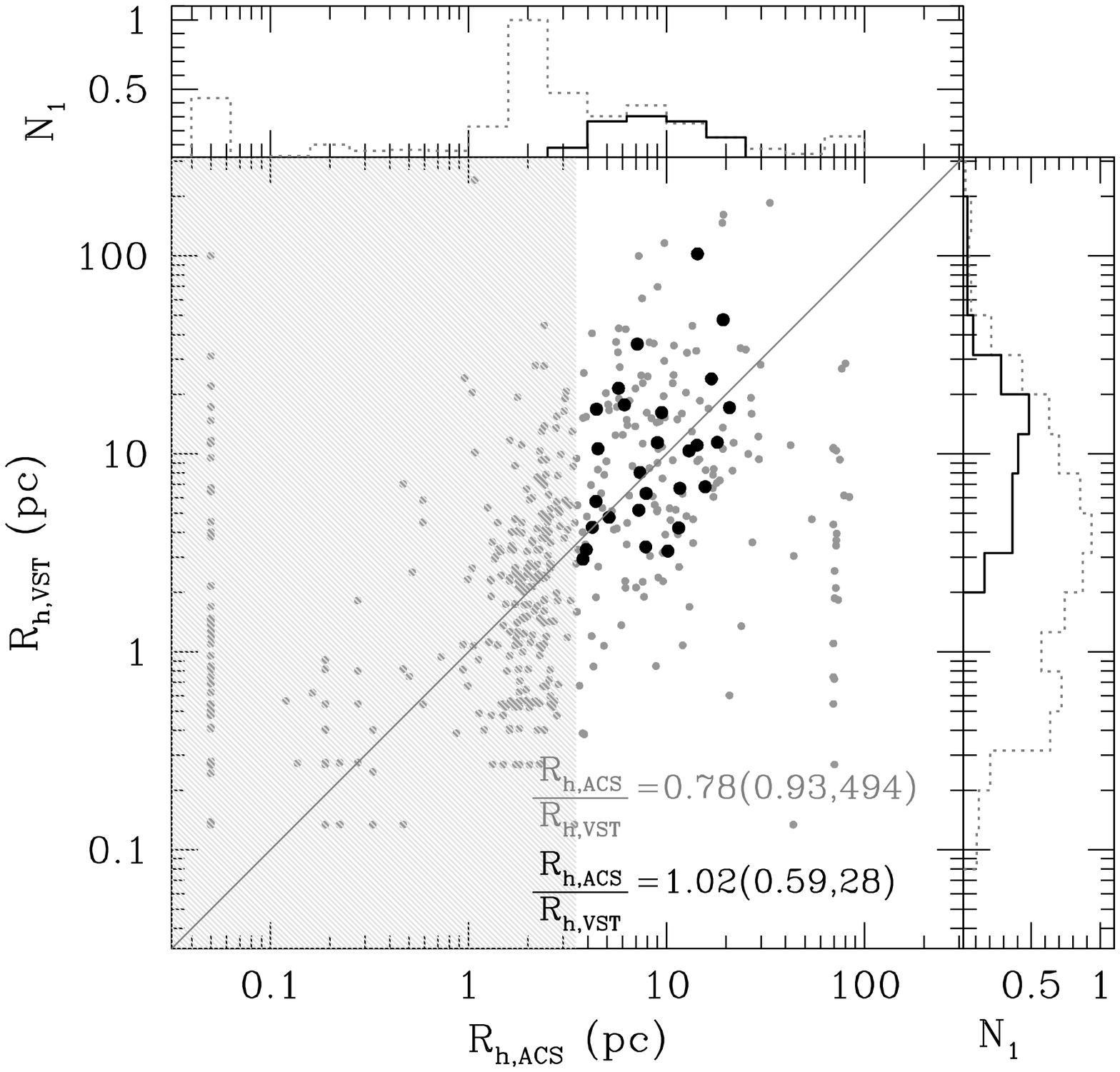} 
  \includegraphics[width=9cm]{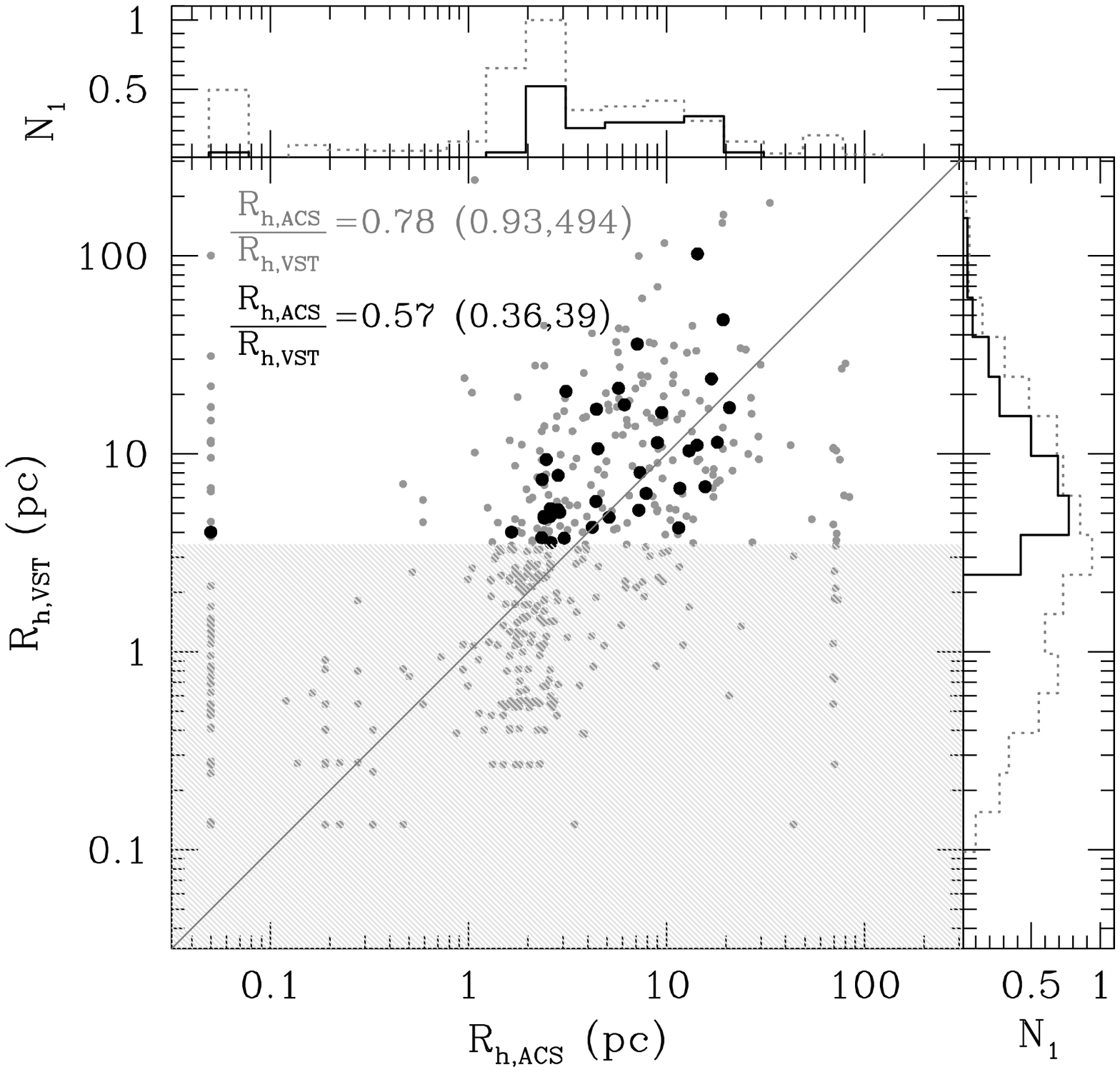} 
  \caption{Size comparison between \sss~ and ACS. Left panels:
    Effective radii for objects common to the ACS and \sss~ catalogs
    (full sample with gray dots, {\it reference} sample with black
    circles). For each sample, the median ratio between the ACS and
    \sss~ $R_h$ is reported, the $rms_{MAD}$ and the number of sources
    used are given in parentheses. The histograms shown in upper and
    right insets use the same colour coding of central panel, and are
    normalized to 1 at peak value for the full sample. The effective
    radii from ACS data are used for the selection (unshaded
    area). The diagonal line represents the 1:1 relation. Right
    panels: as left panels, but the selection is based on $R_h$ from
    \sss.}
   \label{ishape_ref}
   \end{figure*}

\section{Object sizes: comparison with literature and analysis}
\label{sec_sizes}

In this section we present a detailed analysis of the properties of
the SSS in the field of NGC\,3115, and concentrate on the sample of
objects with $R_h$ estimates from \textsf{Ishape}.

\subsection{Comparing \sss~ and HST/ACS derived photometry and sizes}
\label{sec_acs}

As already mentioned, using the photometry and size measurement tools
described in the Section \S\ref{sec_measize}, we ended up with a
catalog containing $\sim$30000 sources. We compared our measurements
with the estimates by \citet{jennings14}, based on ACS/HST $g_{F475W}$
and $z_{F850LP}$ observations. At the distance of the galaxy, all
sources in the field of NGC\,3115 with $R_h\geq 1$ pc appear resolved
at the pixel resolution of the ACS. Thus we assume the $R_h$ measured
from ACS data as best representing the true distribution of $R_h$ for
the GCs in the galaxy, within the limited common area.

The matching of the GCs list from \citet{jennings14} with the VST
catalog (using $0\farcs5$ radius) contains a list of $\sim270$ of
the 360 candidates\footnote{We found a systematic shift in RA, $\Delta
  RA(VST-ACS)\sim+0\farcs6$. In our analysis we applied the
  correction to ACS data.}. Nearly $\sim70\%$ of the unmatched sources
are GCs located in the central galaxy regions, where we do not model
and subtract the galaxy light profile. The remaining $\sim30\%$ of
missing objects are faint sources, undetected in the shallower $i$
image, or objects blended with bright neighbors. The number of missed
sources beyond the central regions drops to $\sim10$ if the sole $g$
band catalog is considered (again faint or blended sources). The
comparisons of $g$-band ACS and \sss~ photometry, shown in Figure
\ref{acsvstmag} versus magnitude and colour, reveals very good
agreement. In the figure, the full ACS catalog (Z. Jennings, private
communication) and the sole GCs sample are considered separately. For
sake of homogeneity the comparison with ACS is done using constant
extinction. The large scatter for the full sample (left panels) is due
to the presence of extended background sources, whose aperture
magnitude does not represent a good estimate for the total magnitude,
neither for ACS nor for \sss. Comparison of photometry for the sole
GCs (right panels in the figure) indicates negligible mean residual
difference, both in magnitude and colour.

We also compared the $R_h$ estimates from \sss~ with the ACS ones. In
the comparison one must note that the list of SSS candidates is not
contaminant free in either catalogs, as it includes background sources
whose (linear) $R_h$ estimates are wrong because they are derived
according to the distance of the galaxy. In spite of this, the ACS
\sss~ comparison is still valid since the same distance modulus is
assumed in both analysis. Moreover, it should also be noted that the
size estimates are not equally good for the full sample of objects
measured. Thus, we define a {\it reference} sample of \sss~ candidates
with reliable structural and photometric parameters, adopting the
following criteria derived on the basis of the comparison with ACS
photometry and objects shape:

\begin{itemize}

\item \textsf{Ishape} signal-to-noise ratio SNR$\geq30$ \citep{larsen99};
\item total relative error on $R_h$ $\leq$30\%;
\item for each source where the iteration to derive $R_h$ was
  successful, \textsf{Ishape} provides a cutout of the image with the
  object analyzed, the model brightness profile, the residuals between
  them, and the weighting map \citep{larsen99}. To reject sources with
  large residuals (see Appendix \ref{app_ucd}, Figure \ref{ucd_bad}),
  but otherwise good SNR and FWHM error, after various experiments
  where we inspected the statistical properties of the residual
  cutouts, we chose a limit of $median/rms\leq0.3$ for good
  candidates. This criterion applies for objects with contaminating
  neighbors or structures not accounted for by the previous criteria;
\item as in Section \S\ref{sec_stat} we adopted colour range $0.4
  \leq(g{-}i)\leq1.25$ mag both for GCs and UCDs;
\item maximum photometric uncertainty $\Delta m_g=0.15$ mag;
\item  axial ratio, $b/a\geq0.3$.
  \citep{vdb84,blake08,cantiello09}.

\end{itemize}

Figure \ref{ishape_ref} shows the \sss~ to ACS size comparison for the full
and reference samples. A first evidence is the ``coma'' shaped
distribution of data. Such behavior highlights the expected lack of
accuracy of \textsf{Ishape} for objects with effective radii below 1/10 the
FWHM, i.e. $R_{h,VST}\lsim3.5$ pc at the distance of the galaxy.

If only sources in the {\it reference} sample and with $R_{h,ACS}\geq
3.5$ pc are used (black filled circles in Figure \ref{ishape_ref},
left panels), the median ratio of ACS and \sss~ $R_h$s is 1.02, while
$rms_{MAD}$ ($rms$ derived from the median absolute deviation) of the
ratio is $\sim0.59$. Thus, for the 29 matched objects in the
{\it reference} sample, the median and standard deviation of the mean are
$1.02\pm0.11$, providing a satisfactory agreement for the ACS and \sss~
samples when limited to the {\it reference} sample. Nevertheless, we
must highlight that the $R_h$ estimates for single objects can differ
up to a factor $\sim$5 even for $R_{h,ACS}\geq 3.5$ pc. Further
details on the differences between ACS and \sss~ size estimates for
extended objects are given in Appendix \ref{app_ucd}.

However, for the typical VEGAS target the selection will only rely on
$R_h$ measurements from VST images. Right panels of Figure
\ref{ishape_ref} show the same data of left panels, using the
$R_{h,VST}$ values for the selection instead of $R_{h,ACS}$. The ACS
to \sss~ comparison worsens, as the $R_{h,ACS}$ to $R_{h,VST}$ mean
ratio and standard deviation of the mean are 0.57$\pm$0.06. It is
interesting to note that, taking as lower limit $R_{h,VST}=2$ pc, we
obtain $\frac{R_{h,ACS}}{R_{h,VST}}=0.78\pm0.08$ ($rms=0.55$, 46
objects). This suggests that even though the nominal limit for \textsf{Ishape}
is 1/10 the FWHM, or $\sim3.5$ pc at the distance of NGC\,3115, the
tool allows to separate stars from extended sources down to 2 pc.

%____ISHAPE TESTS ACS COMPARISON__________________________________________ 
   \begin{figure}[h]
   \centering
\resizebox{\hsize}{!}{\includegraphics{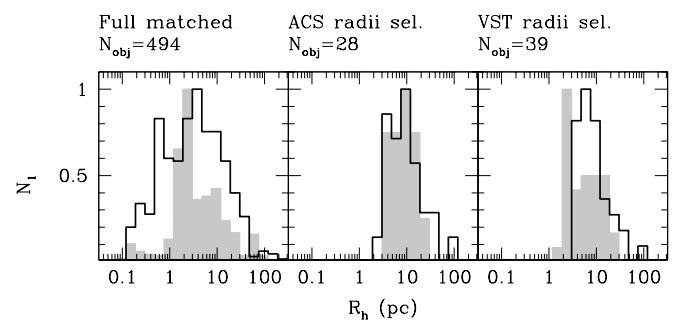}}
  \caption{$R_h$ distribution for ACS and \sss~ samples. From left to
    right: $R_h$ distribution for the full list of matched ACS and
    \sss~ sources, for the sample selected using $R_{h,ACS}$, and for
    the sample selected using $R_{h,VST}$ (see text). Shaded
    histograms refer to $R_{h,ACS}$ distributions, solid thick lines
    to $R_{h,VST}$. All histograms are normalized to one at peak
    value.}
   \label{2select2}
   \end{figure}

To further inspect the issue, Figure \ref{2select2} shows the $R_h$
distributions for: $a)$ the full list of matched sources from ACS and
\sss~ (left panel, ACS data in gray, VST data with thick black line),
$b)$ objects in the {\it reference} sample selected using
$R_{h,ACS}\geq3.5$ pc (middle panel), and $c)$ objects in the {\it
  reference} sample selected using $R_{h,VST}\geq3.5$ pc (right
panel). The ACS and \sss~ distributions appear quite similar in middle
panel (case $b$). The $R_h$ distributions based on \sss~ half-light
radii (right panel, case $c)$ shows a shift, with \sss~ radii being on
average larger, and missing the peak at $R_h\sim2$ pc seen in the ACS
data. This behavior is due to the sources more compact than 3.5 pc,
which are scattered over the entire 3.5-20 pc interval when $R_h$
estimates from VST are used. In particular, the list of common SSS
candidates in the {\it reference} sample goes from 28, with the
selection based on the ACS radii, to 39 with the $R_h$ selection from
VST data. On such basis, we estimate that for values of $R_h\geq3.5$
pc the {\it reference} sample contains $\sim30\%$ objects with
unreliable effective radii, spread over the entire $R_h$
distribution. Needless to say that the sample of matched objects with
high quality \sss~ sizes is relatively small (28 or 39, depending on
the selection), thus making hard to generalize the results of the
comparison over the entire VST field of view.

Finally, with the aim of deriving a catalog of GC candidates from the
sole VST data, and estimating the contamination taking as {\it
  reference} the ACS GCs list, we carried out the following blind
test. We adopted the selection criteria given at the beginning of this
section, with the additional requirement that GC candidates must have
$2\leq R_h\leq 8$ pc \citep[we adopted the same $R_h$ used for GC
  by][for ACS data]{jennings14}, and matched such \sss~ list to the
sample of GC candidates from ACS. The results is that $\sim20\%$ of
the candidates ($\sim10$ over $\sim50$) are not present in the GC list
from ACS\footnote{The result is not much sensitive to the particular
  choices of \textsf{Ishape} input parameters (see Appendix
  \ref{app_ishape}, Figure \ref{ishape_acsvst_stat}).}. In contrast,
adopting as lower limit $R_h=3.5$ pc, the number of matching sources
is $\sim30$ and the contamination is nearly doubled. In other words,
the test points out that the results from \textsf{Ishape} allow to
distinguish between compact and extended sources down to $R_h=2$ pc,
although the exact value of the effective radius is reliable only
above $\sim3.5$ pc.

 %____________________________________
  \begin{figure*} 
   \centering
\includegraphics[width=9cm]{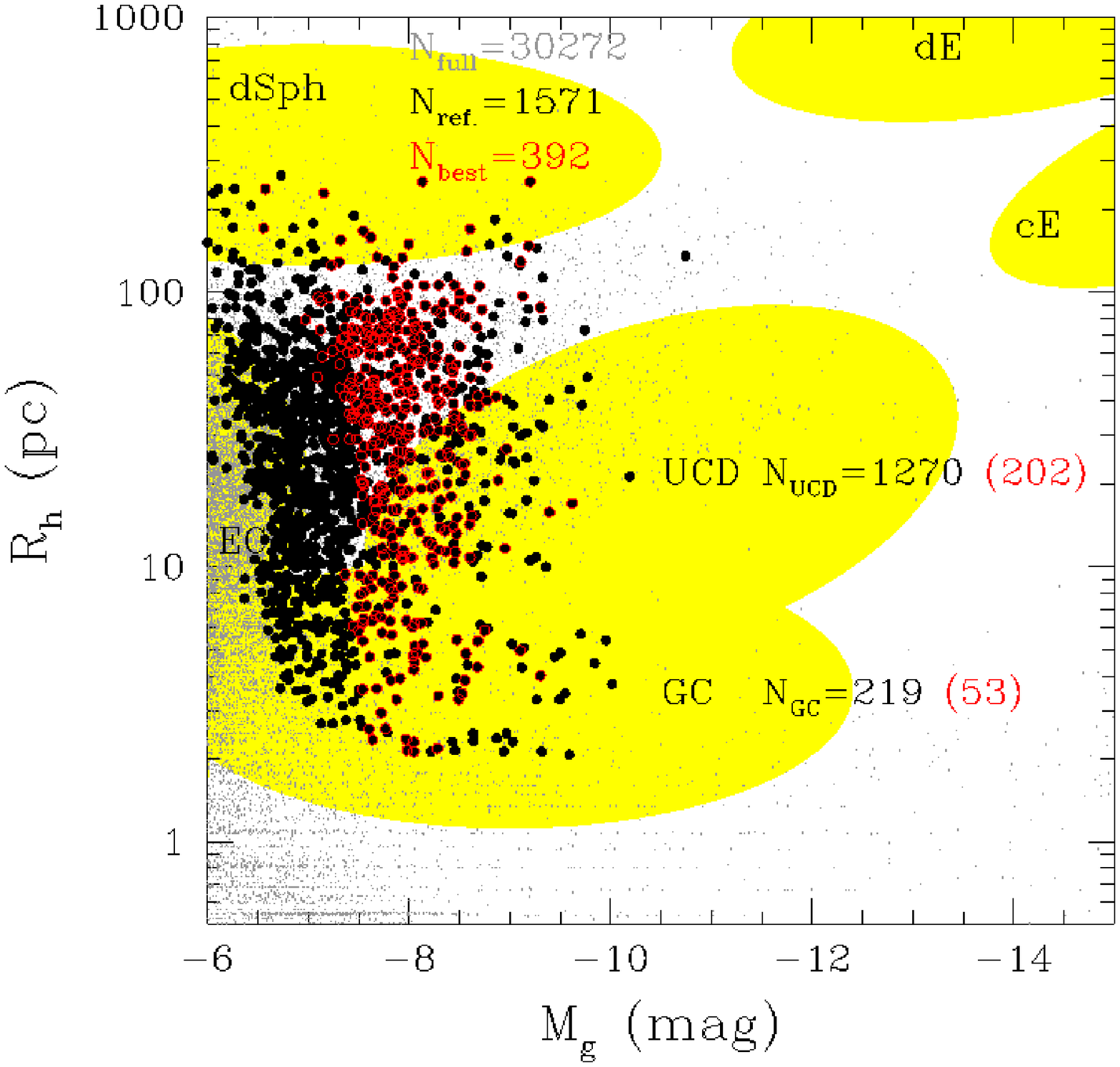}
\includegraphics[width=9cm]{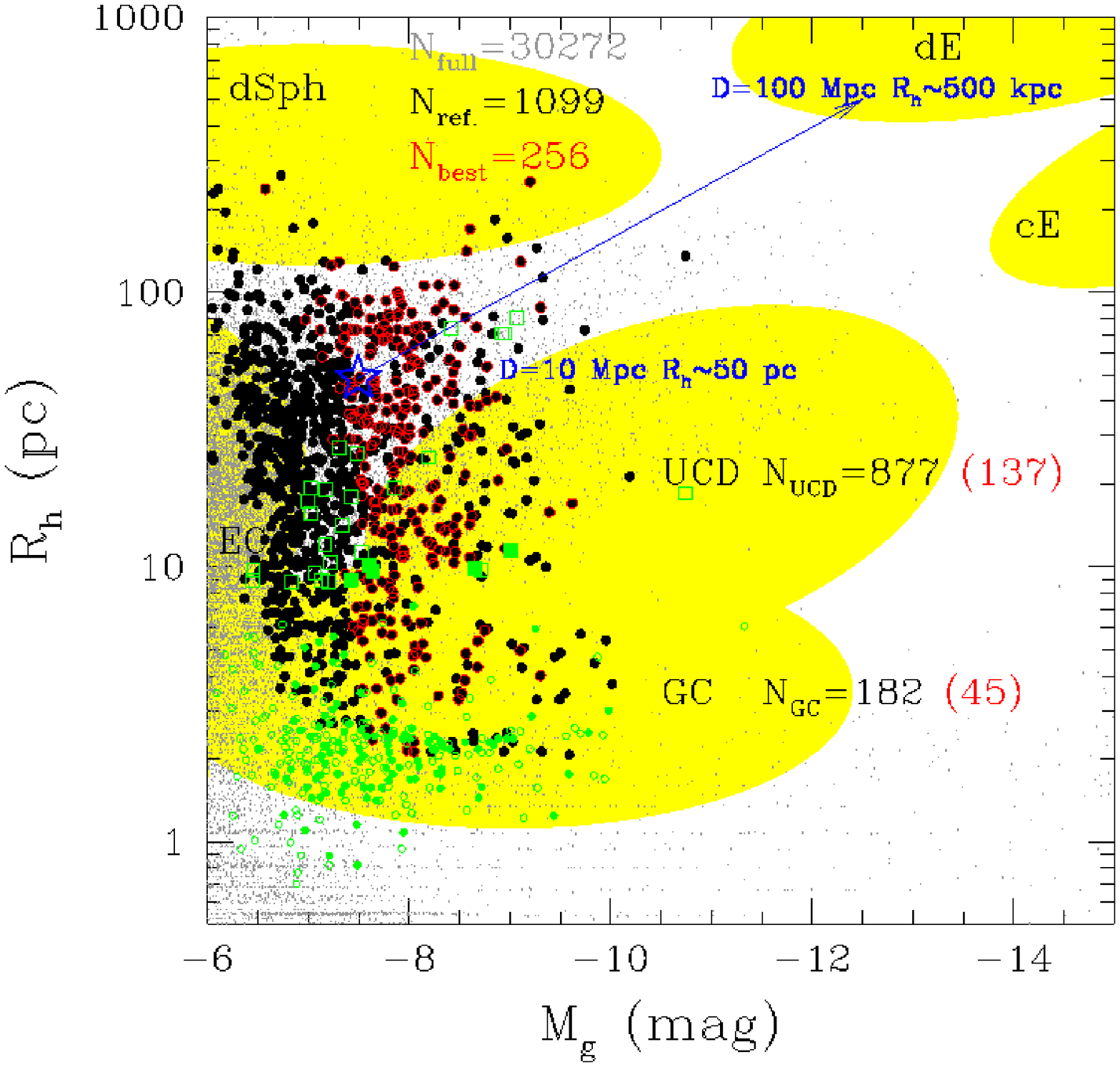}
  \caption{Magnitude-size diagram. Left panel: magnitude versus
    effective radius for the \sss~ field centered on NGC\,3115. Colour
    coding for black and gray dots is as in previous figures, with the
    addition of the best sample selection (red empty circles). Yellow
    regions show the {\it mean} loci of the labeled SSS classes. The
    number of UCD and GC candidates, $N_{UCD}$ and $N_{GC}$, for the
    {\it reference} and {\it best sample} (given in parenthesis) are
    also reported. Right panel: as left panel but only sources with
    $R_{gal}\leq R_{bg}$ are plotted. The blue arrow shows the
    direction the points are shifted if the object lies at larger
    distance. We included the ACS sample using green symbols: GCs
    shown with circles, UCDs with squares. For the ACS sample,
    spectroscopically confirmed GCs from \citet{arnold11} are plotted
    as solid symbols.}
   \label{mgreff5}
   \end{figure*}

In conclusion, using the {\it reference} sample obtained from the
coupling of photometric and spatial extent properties, the present
\sss~ catalog can be used to: \\ $i)$ obtain a list of GC candidates,
selected in the range $2\leq R_h (pc) \leq 8$, with an expected
contamination of $\sim20\%$, poorly populated because of the narrow
selection adopted. The number of GC candidates over the entire area
inspected, $\sim52.5\arcmin\times52.5\arcmin$ (or $\sim145$ kpc
$\times 145$ kpc), selected on the given photometric and size criteria
is $N_{GC}\sim220$. However, {\it for radii below 3.5 pc, the $R_h$
  are only used as an effective binary selection criterion}
(i.e. $R_h\geq2$ ($<2$) meaning extended (point-like) source), as this
limit is smaller than the nominal limit of the tool; \\ $ii)$ a
catalog of extended objects with $R_h\geq3.5$ pc, having a
contamination of $\sim30\%$ objects with unreliable $R_h$ estimates.

\subsection{GC and UCD population properties based on colour and size selection criteria}
\label{sec_best}

Figure \ref{mgreff5} shows the size versus magnitude diagram obtained
using the {\it reference} sample, i.e. with the selection criteria
described in the previous section, and adopting for all objects in the
field the same distance modulus. In the right panel, we plotted only
SSSs at galactocentric distance $R_{gal}\leq R_{bg}=23\arcmin$,
corresponding to $\sim65$ kpc at the distance of NGC\,3115. The
approximate regions for GCs, UCDs, extended clusters (ECs), dwarf
spheroidal (dSphs), dwarf ellipticals (dEs) and compact ellipticals
(cEs) are shown and labeled \citep[yellow regions; mean loci are taken
  from][]{brodie11,bruens12}.

From Figure \ref{mgreff5} (left panel), we find that a large fraction
of selected objects in the {\it reference} sample falls within the
avoidance region at $M_g\sim-7.5$ mag and $R_h\gsim50$ pc
\citep{forbes13}. The situation does not seem to improve much even if
only sources in the {\it reference} sample and with SNR$\geq60$ ({\it
  best sample} hereafter, red circles in the figure) are
considered. On the other hand, the number of sources in the avoidance
area is lowered if only sources within $R_{gal}\leq R_{bg}$ are taken
(Figure \ref{mgreff5}, right panel), but still significantly
large. Moreover, we must highlight the large number of UCD candidates
even for the {\it best sample} and for $R_{gal}\leq R_{bg}$ sources
($N_{UCD}=137$).

In spite of the results by \citet{forbes13}, who find that the
avoidance zone is the result of a selection bias and confirmed the
presence of various SSSs within the region, a large fraction of
sources in the avoidance area are likely background galaxies (some of
which are recognizable by eye). As shown by the arrow in Figure
\ref{mgreff5}, in fact, a background source should move toward larger
absolute radii and brighter when larger distance moduli are
considered.

In right panel of Figure \ref{mgreff5}, we also plotted the GCs (green
circles) and UCDs (squares) data from \citet{jennings14}. The
spectroscopically confirmed GCs and UCDs from \citet{arnold11} are plotted as
solid symbols. Two interesting elements here are $i)$ the nice overlap
of the overdensity region for spectroscopically confirmed GCs and \sss~
selections, at $-7.5\leq M_g\leq-9$ and $R_h\sim2-2.5$ pc, and $b)$
the presence of UCDs from the ACS sample outside the region where they
typically occur. Three UCD candidates from the ACS sample, in fact,
lie at $M_g\sim-9$ mag and $R_h\sim 80$ pc, i.e. within the zone of
avoidance (if any). Two other UCDs have $M_g\sim-6.5$ mag and $R_h\sim
10$ pc, typically associated with the EC region (see also the discussion
in Appendix \ref{app_ucd}). This clearly shows that the distinction
between the different SSS types is not trivial, and sometimes contains
elements of arbitrariness.

\begin{figure}[ht]
   \centering
 \resizebox{\hsize}{!}{\includegraphics{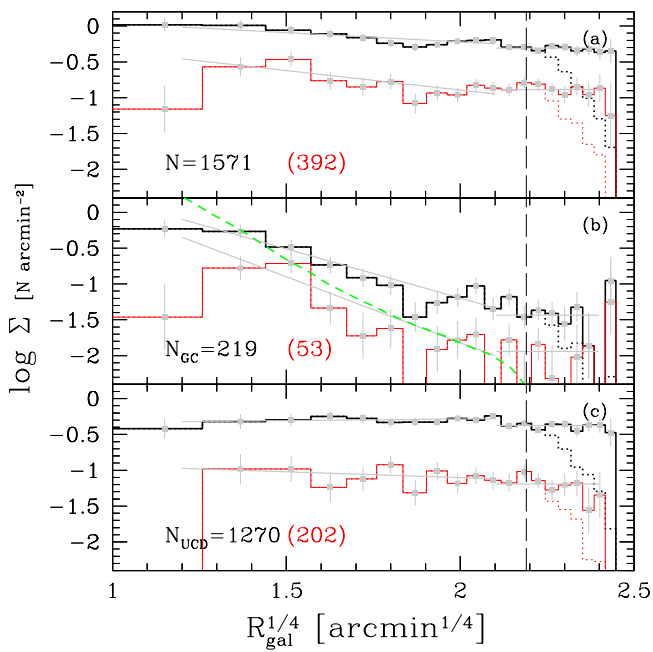}}
  \caption{Radial surface density distribution of sources in the {\it
      reference} (black histograms) and best (red histogram)
    samples. Panel $(a)$: all $R_h$ values are taken. The $r^{1/4}$
    fit to data is shown with gray lines. The dotted lines show the
    histograms with no correction for areal coverage. The vertical
    long-dashed line is the position of limiting galactocentric radius
    $R_{bg}$. Panel $(b)$: as upper, but only GC candidates are
    plotted. The green-dashed line shows the galaxy surface brightness
    \citep[from][$g$ band, arbitrary scale]{capaccioli14}. Panel
    $(c)$: as panel $(a)$, but for UCD candidates.}
   \label{histocap5}
   \end{figure}

To further inspect the issue, we analyzed the surface density
distribution of sources versus galactocentric radius, and versus
$R_h$.  Figure \ref{histocap5} shows the radial surface density
distribution for the {\it reference} and {\it best sample}s (black and
red histograms in panel $(a)$, respectively), for GC candidates
($2\leq R_h~(pc)\leq8$, panel $(b)$), and for UCD candidates
($8<R_h~(pc)\leq100$, panel $(c)$)\footnote{To estimate the density of
  sources, the effective area coverage in each annulus is corrected
  for the annular area outside the image and for the central
  uninspected regions (dashed histograms show the uncorrected
  distributions). Poisson statistics is adopted to estimate the
  errors.}. The gray lines in the figure show the $r^{1/4}$ profile
assuming a constant background, obtained from the flat region at
galactocentric radius $R_{gal}>R_{bg}$. In each panel we also report
the total number of objects selected for the {\it reference} and {\it
  best sample} (the latter in parentheses). The surface density for
the sample with no selection on sizes (panel $(a)$) shows an obvious
correlation with $R_{gal}$, and a flattening at $R_{gal}\geq R_{bg}$,
suggesting that sources beyond this radius are most likely background
galaxies or foreground stars. The radial density profile for GC
candidates follows a de Vaucouleurs density profile out to $R_{bg}$,
as for the galaxy light, providing further proof to the actual
membership of the objects selected to the GCs population, and
supporting the role of the object-size analysis carried out
here. However, one must not neglect the presence of a fraction of
background sources. For what concerns the distribution of objects with
UCD-like radii, the {\it reference} sample does not show a tendency
for a radial trend. The result is not surprising, given the small
number of expected UCD-candidates and the large fraction of
contaminants. It is noteworthy, though, that the UCDs in the {\it best
  sample} show hints of a radial trend.

By integrating the fitted de Vaucouleurs $r^{1/4}$ density profiles,
from zero to $R_{bg}$ for both GC and UCD density profiles, after
subtracting the total number of background sources we find for the
{\it reference} ({\it best}) GC sample $N_{GC}\sim113$ ($\sim42$), and
for the {\it best} UCD sample $N_{UCD}\sim30$. The comparison of these
numbers, in particular for the GCs, with the numbers in Figure
\ref{mgreff5} (right panel) confirm our previous results that the
contamination for the reference GC sample is $\sim30\%$, and also
indicates that the {\it best sample} suffers from very small
contamination. The numbers are quite different for UCDs, given the
higher confusion with extended background sources. In such case, the
coupling with data in other passbands will greatly reduce the
contamination.
 
\begin{figure}[ht]
   \centering
  \resizebox{\hsize}{!}{\includegraphics{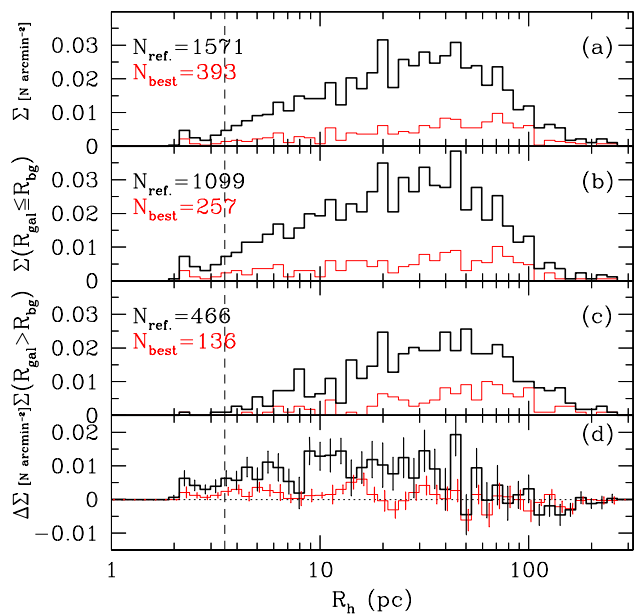}}
  \caption{$R_h$ surface density distribution for the {\it reference}
    sample (black lines) and for the {\it best sample} (in red,
    $\Sigma$ in units of number of objects per square arcminute). The
    panels, from upper to lower, show the surface density over the
    full inspected frame, for sources at $R_{gal}\leq R_{bg}$, for the
    background area at $R_{gal}> R_{bg}$, and for the difference
    between the inner and outer 23$\arcmin$, respectively. The
    vertical dashed line marks the 3.5 pc limit.}
   \label{mgreff_histo2}
   \end{figure}

Figure \ref{mgreff_histo2} shows the $R_h$ distributions for the {\it
  reference} and {\it best} samples, normalized to the area inspected:
full detector area, objects within the $R_{gal}\leq R_{bg}$ area,
objects outside $R_{gal}>R_{bg}$, and the difference between latters
(from panel $(a)$ to panel $(d)$, respectively).

The {\it reference} sample over the entire VST area (upper panel),
shows slightly increasing surface density for increasing $R_h$ up to
$\sim$50 pc. While the $R_h$ distribution for the {\it best sample} is
rather flat. The differences of the surface density for objects within
$R_{bg}$ or in the outside area, shown in panels $(b)$ and $(c)$ of
Figure \ref{mgreff_histo2}, are quite obvious, especially for the
over-density of objects with GC-like radii, $3.5 \leq R_h~(pc) \leq
8$. Because of the contamination, the density of sources with GC-like
radii is non-zero in the outer radius (panel $c$). This is in part due
to the expected fraction of galaxy GCs that might lie at large
galactocentric distances (see Section \ref{sec_stat}), while most of
the contribution comes from background contamination. Indeed, the mean
density of objects with $3.5 \leq R_h~(pc) \leq 8$ at radii
$R_{gal}>R_{bg}$ (panel $c$) is $\sim$30\% the one at $R_{gal}\leq
R_{bg}$ (panel $(b)$). This result confirms our previous estimate of
the fraction of contamination for the colour and size selected {\it
  reference} sample. Furthermore, we note that the density in the
background region is $\lsim$15\% the one in the inner regions for the
{\it best sample}.

 %____________________________________ 

The differences between inner and outer density are better seen in
panel $(d)$ of Figure \ref{mgreff_histo2}, where $\Delta
\Sigma\equiv\Sigma(R_{gal}\leq R_{bg})-\Sigma(R_{gal}>R_{bg})$ versus
$R_h$ is shown. Here, the surface density of sources with $R_h\gsim70$
pc is consistent with zero. More in details, the $\Delta \Sigma$
distribution for the {\it reference} sample (black histogram), is
generally consistent with zero density from $R_h\gsim$50 pc (with some
possible candidates at $R_h\sim55-60$ pc), and for $R_h\sim8$ pc,
while for the {\it best sample} various regions are compatible with
zero density (e.g $R_h\sim7-10$ pc, $\sim20-30$ pc, and $\geq40$ pc).
These results, imply that the surface density of objects with such
$R_h$ values is constant over the inspected area, as would be expected
from a uniform background contamination. In other words, panel $(d)$
suggests that the majority of sources falling in the zone of avoidance
(Figure \ref{mgreff5}) are background galaxies. Second, the
over-abundance of sources with $R_h$ having GCs-like radii appears
clearly both for the {\it reference} and {\it best
  samples}. Furthermore, for the {\it reference} sample, we find a
positive density of sources around the characteristic $R_h$ values of
UCDs (between 10 and 40 pc), which confirms the membership to this
class for some of the selected objects. Such over-density, though, is
weaker for the {\it best sample}, and possibly consistent with zero in
some cases.

\section{Summary}

In this paper we presented the first results of the VEGAS survey for
the specific science case of small stellar systems, SSSs. We described
the methodology for the photometry and the size analysis of SSS
candidates in the field of NGC\,3115, a well studied lenticular
galaxy, and showed the potential of the survey in providing original
results on SSS-related science.

The VEGAS survey will collect the deep $g$ and $i$ imaging of bright
ellipticals in the Southern hemisphere, possibly complemented with $r$
for most of the targets, and also with $u$ band observations for
selected galaxies. One of the great advantages of \sss~ is the use of
wide field imaging, $\sim1$ square degree, which allows to study the
properties of SSSs out to very large galactocentric distances, with an
accurate characterization of the background contaminating objects. For
the specific case of NGC\,3115 we inspected the properties of SSSs out
to $\sim23\arcmin$, i.e. more than twenty times the effective radius
of the galaxy.

We first analyzed the properties of the GCs system. Being the
population of SSSs numerically most abundant in the galaxy, GCs
properties can be derived using the sole photometric information,
colour and magnitudes, by comparing the surface density of sources in
the inner galaxy regions with the density in the outer regions. Our
results can be roughly divided into two groups: $i)$ results that
repeat previous analysis, giving us the chance to confirm the
reliability of this study; $ii)$ new results allowed by the use of the
wide-field imaging. In the first group we include:

\begin{itemize}
\item the GCs have a bimodal \gi distribution with peaks at $\sim0.75$
  and $\sim1.0$ mag;

\item red GCs are more centrally concentrated than blue GCs;

\item as for the galaxy light, the radial density of GCs follows a de
  Vaucouleurs $r^{1/4}$ profile, but with a shallower slope.

\item the turnover magnitude of the $g$-band GCLF, $M^{TOM}_g$,
  coupled with the calibration from the ACSVCS survey, implies a
  distance modulus $\mu_0=29.95\pm0.3$ in good agreement with the
  literature.
\end{itemize}

Such achievements support the results of previous studies, some of
which carried out with 8-10m class telescopes, and are further
complemented by the following compelling results:

\begin{itemize}
\item the colour bimodality extends to more than $\sim20$ galaxy
  effective radii;

\item the blue GCs show a tendency towards bluer colour at larger
  galactocentric radii $R_{gal}$, while red GCs seem to have a nearly
  constant colour with $R_{gal}$;

\item the galaxy light has a steeper density profile than the GCs,
  whether the blue or total fractions of GCs is taken into account;

\item the slope of the surface density profile for red GCs at
  $R_{gal}\geq7.5\arcmin$ matches with that for the galaxy light,
  while a red GCs overdensity appears in the inner galaxy regions;

\item the ratio of blue to red clusters shows a trend with $R_{gal}$,
  with the fraction of blue GCs being slightly larger at larger radii;

\item by analyzing separately the blue and red GCs we find a $\Delta
  m_g^{TOM}\sim0.2$ mag, with the blue TOM being brighter; 

\item we do not find an obvious dependence of $M_g^{TOM}$ with
  $R_{gal}$.

\end{itemize}

Both the colour and luminosity properties obtained are consistent with
similar existing studies of the GC system in other early-type
galaxies.

The presence of a bimodal GC system, with blue GCs more extended than
the galaxy stellar light, and a deficiency of red GCs in the inner
regions, have already been observed in other early-type galaxies
brighter than NGC\,3115, and support a scenario where blue GCs are
associated with the galaxy halo, while red ones are more centrally
concentrated and associated with the bulge stellar component in the
galaxy. The overall observed properties might suggest that the galaxy has
undergone a relatively quiescent evolution, without major star-forming
events.   
\\
\\
\noindent Adding the spatial extent of the sources to the colour
information gives a further criterion for selecting SSSs, in particular
GCs and UCDs. We used \textsf{Ishape} to determine the effective
radius $R_h$ of slightly extended objects in the field. By comparing
our estimates for the objects in the {\it reference} sample with the
ones in the literature, obtained from ACS data, we find on average
satisfactory agreement. However, the $R_h$ estimates for single
objects can differ up to a factor $\sim$5 between ACS and
\sss. Furthermore, the result is sensitive to the $R_h$ estimate taken
as reference (ACS or \sss), because of the contamination of the \sss~
sample. The various comparisons with the literature and with
inner/outer galaxy regions suggest that the level of fore/background
contamination of our {\it reference} sample is $\sim$ 30\%, possibly
reduced to one half for the (poorer) {\it best sample}.  Future
studies with new \sss~ $u$ and $r$ band data will be used to further
constrain the properties of other, less populated classes of SSSs,
like cEs, in the field of NGC\,3115.

In spite of the large uncertainties posed by the estimate of $R_h$,
the results obtained are encouraging, suggesting that similar analysis
could be successfully carried out for the other targets in the
survey. Although, at larger distances, the study of sizes will be
limited to the most extended SSSs (UCDs, cEs), excluding the GCs
component for most of the targets beyond $\sim10$ Mpc distance.
\\ \\  The results of this work, on one hand confirm the existing
  studies, thus support the validity of the analysis scheme developed
  here using data from the 2.6m VST telescope. On the other hand they
  provide new and independent results - especially for what concerns
  the GCs properties out to the previously unreached galactocentric
  distance of $\sim65$ kpc - showing the great potential for future
  applications to other VEGAS targets, in particular for the part of
  the sky not accessible to similar facilities.  \\ \\
\noindent 
As a final remark, we highlight that, at survey completion, for
  most of the VEGAS targets observations in at least one more
passband other than $g$ and $i$ will be available. The selection of
SSSs with a further optical color would certainly reduce the percentage
of contaminants, especially if $u$ band photometry is
included. However, a contamination free catalog based on purely
optical photometry is basically unattainable. Since the coupling of
optical data with just one near-IR band is very effective in reducing
the fraction of contaminants to the GC and UCD populations to
$\lsim5$\%, the \sss~ catalogs will be perfectly suited to be
complemented with single-band near-IR imaging (e.g. with a large
format near-IR imager like VISTA), to define the most complete and
clean SSS catalogs possible, essential for, e.g., future spectroscopic
follow-up.

\begin{acknowledgements}

The optical imaging is collected at the VLT Survey Telescope using the
Italian INAF Guaranteed Time Observations. The data reduction for this
work was carried out with the computational infrastructures of the VST
Center at Naples (VSTceN). We gratefully acknowledge INAF for
financial support to the VSTceN. Part of this work was supported by
PRIN-INAF 2011 (P.I.: G. Marconi), FIRB-MIUR 2008 (P.I. G. Imbriani),
PRIN-INAF 2011 (P.I.: A. Grado). M.P. acknowledges finical support
from project FARO 2011 from the University of Naples Federico
II. D.A.F.  thanks the ARC for financial support via DP130100388.
We are grateful to John P. Blakeslee, and Zach
Jennings for useful discussions related to this work.

This research has made use of the NASA/IPAC Extragalactic Data-base
(NED) which is operated by the Jet Propulsion Laboratory, California
Institute of Technology, under contract with the National Aeronautics
and Space Administration. This research has also made use of the
SIMBAD database, operated at CDS, Strasbourg, France, and of the
HyperLeda database (http://leda.univ-lyon1.fr).

\end{acknowledgements}

%%%%%%%%%%%%%%%%%%%%%%%%%%%%%%%%%%%%%%%%%%%%%%%%%%%%%%%
%%%%%%%%%%%BIBLIOGRPHY %%%%%%%%%%%%%%%%%%%%%%%%%%%%%%%%
%%%%%%%%%%%%%%%%%%%%%%%%%%%%%%%%%%%%%%%%%%%%%%%%%%%%%%%

%%%%%%%%%%%%%%%%%%%%%%%%%%%%%%%%%%%%%%%%%%%%%%%%%%%%%%%
%%%%%%%%%%%TABLES %%%%%%%%%%%%%%%%%%%%%%%%%%%%%%%%%%%%%
%%%%%%%%%%%%%%%%%%%%%%%%%%%%%%%%%%%%%%%%%%%%%%%%%%%%%%%

%%%%%%%%%%%%%%%%%%%%%%%%%%%%%%%%%%%%%%%%%%%%%%%%%%%%%%%%%%%%%%
%%%%%%%%%%%%figures%%%%%%%%%%%%%%%%%%%%%%%%%%%%%%%%%%%%%%%%%%%
%%%%%%%%%%%%%%%%%%%%%%%%%%%%%%%%%%%%%%%%%%%%%%%%%%%%%%%%%%%%%%

\begin{appendix}

\section{On the completeness and the edge detection functions}
\label{appendix_cc}

\subsection{Completeness correction}
\begin{figure} 
   \centering
  \resizebox{\hsize}{!}{\includegraphics{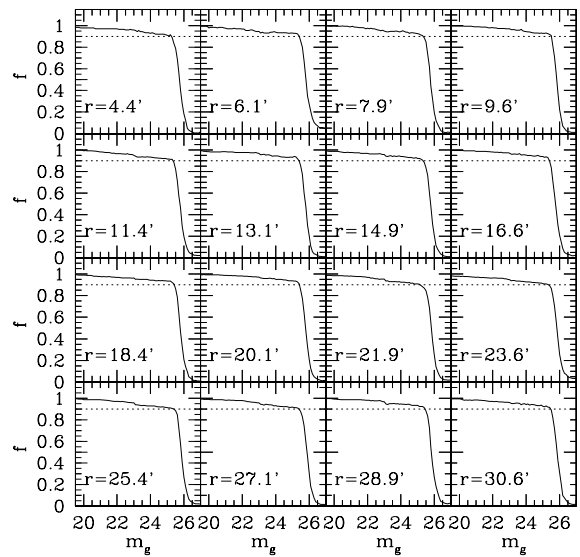}}
  \caption{Completeness function estimated at different galactocentric
    radii. The label refers to the mean radius of the annulus in
    arcminutes. The dotted lines show the 90\% completeness limit.}
   \label{plotcf3}
   \end{figure}

The completeness function of the $g$-band images was determined by
adding artificial stars to the original images and then reprocessing
them as described in Section \ref{sec_photo}. The ratio between the
number of artificial stars added, and the number of stars recovered
provides the estimate of the completeness. We added stars using a grid
pattern, with $\sim20\arcsec$ increments in x and y. Since the field
is dominated by the light from NGC\,3115, the correction for magnitude
completeness depends on the angular distance from the galaxy center
\citep[e.g.][]{cantiello07c}. The radial dependent completeness
function is shown in Figure \ref{plotcf3}. To correct the luminosity
functions the number of objects at given magnitude is multiplied by
$1/f$ using the proper function at each galactocentric distance.

 %______________________________________ 

\subsection{Edge detection filter}

%______________________________________ 
  \begin{figure} 
\resizebox{\hsize}{!}{\includegraphics[]{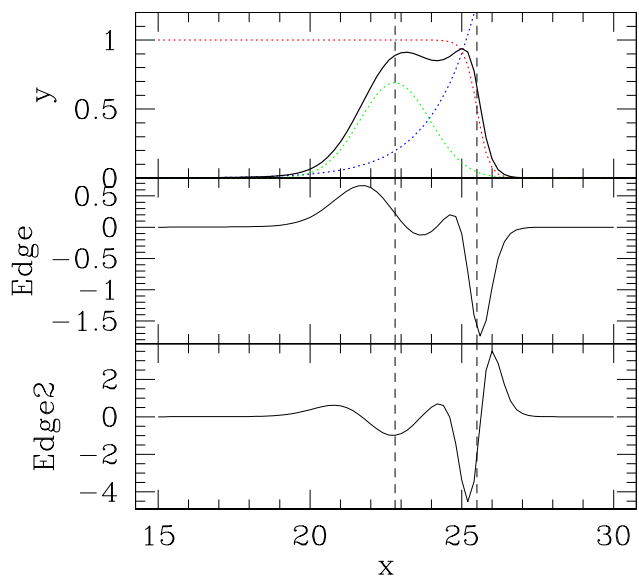}}
  \caption{Upper panel: analytic GCLF (green line), background
    galaxies (blue), and completeness function (red) are combined in
    the total expected total luminosity function (black solid
    line). The GCLF peak (at x=22.8), and 50\% completeness limit (at
    x=25.5) are shown with dashed vertical lines. Middle panel: edge
    detection function applied to the analytical formula of the total
    luminosity function. Lower panel: second-run edge detection.}
   \label{simuledge}
   \end{figure}

In Figure \ref{simuledge} we analyze the behavior of the
edge-detection function on a composite function similar to the one
expected for the sources in the field of NGC\,3115. The function
inspected is the sum of a gaussian GCLF \citep{harris01} and a power
law for background galaxies \citep{tyson88,bernstein02}, times a
completeness smoothed step function (green, blue and red line,
respectively). The edge-detection function, in a first approximation,
is a derivative function and shows an inflection point at the GCLF
turnover magnitude (Figure \ref{simuledge}, middle panel). A second
run of the edge function - Edge2, roughly a second derivative -
reaches a local extrema at the TOM. Thus, in first approximation, the
turnover of the GCLF can be found in correspondence of an inflection
and a local extrema in the Edge and Edge2 functions, respectively.

\section{Some details on \textsf{Ishape}}
\label{app_ishape}
\textsf{Ishape} uses a PSF subsampled by a factor 10 relative to the
resolution of the science image. To model the PSF we used the DAOPHOT
package within IRAF and, to reduce the chance of contaminating the PSF
modeling with GCs in the galaxy, we included in the list of PSF
candidates unsaturated sources with $g{-}i\leq0.3$, $g{-}i\geq1.7$ and
$m_g\geq 18$ mag. To account for PSF variations across the image, we
set DAOPHOT VARORDER=2, which means that the PSF is quadratically
variable over the image. Then, the frame was divided in a grid of
$5\times5$ equal subframes, and the model PSF for \textsf{Ishape}
evaluated in the center of each subframe.

Within \textsf{Ishape}, we adopted the ``KING30'' profile, i.e. the
\citet{king62} model with concentration parameter $c=30$, which is
typical for marginally resolved GCs and UCDs \citep{larsen00,blake08}.

\begin{table}
\tiny
\caption{Main Ishape parameters used for the analysis.}
\centering
\begin{tabular}{l c c}
\hline\hline
Parameter        &   Value      &       Explanation                                   \\
\hline
PSF             & Moffat25      & Input PSF from DAOPHOT \\
\hline 
FITRAD          &      12       &   Fitting radius \\
CENTERRAD       &      3        &   Maximum centering radius \\
CLEANRAD        &      3        &  Cleaning radius  \\
CTRESH          &      2        &  Threshold for cleaning \\         
MAXCITER        &      5        &  Maximum number of iterations \\
CENTERMETHOD    &      MAX        &     Centering method \\
SHAPE           &      KING30     &  Shape used for profile fitting \\
FWHMMAX         &      20.0       & Maximum FWHM \\
ITMAX           &      200        & Maximum number of iterations \\
ELLIPTICAL      &      YES        & Use elliptical model \\
EPADU           &      11.5       & $e-$/ADU conversion factor \\
RON             &      7.0        & CCD read-out noise \\
CALCERR         &      YES        &  Calculate errors \\
\end{tabular}
\label{tab_ishape}
\end{table}

\begin{figure}
   \centering
\resizebox{\hsize}{!}{\includegraphics{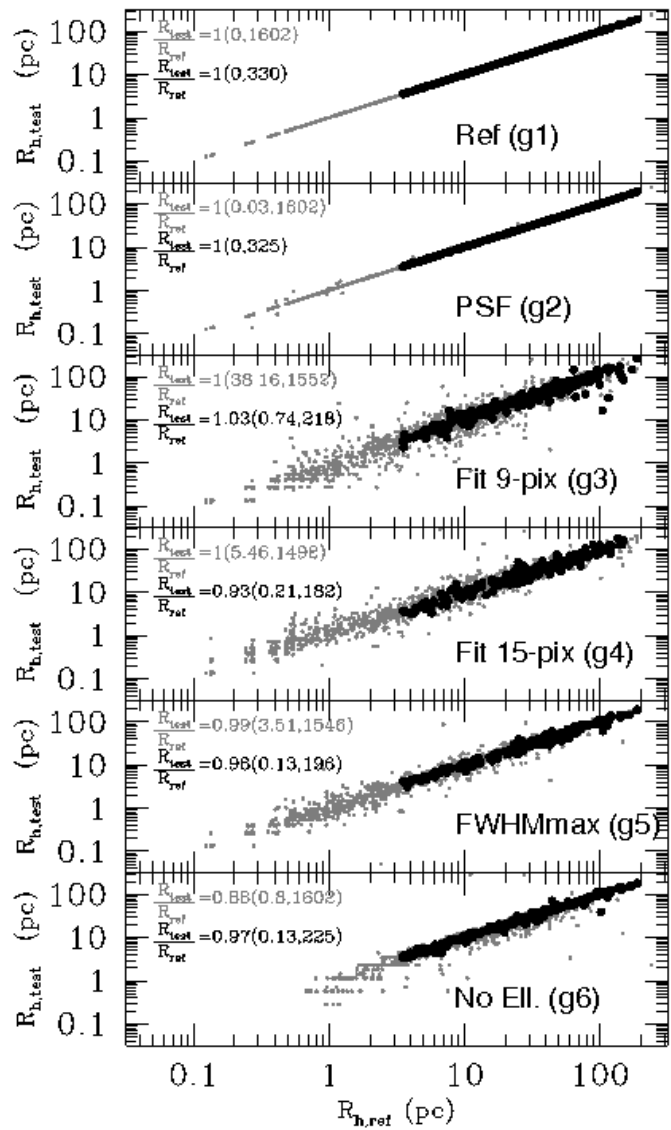}}
   \caption{Comparison of effective radii with different \textsf{Ishape} input
     parameters. Grey circles refer to the full list of matched
     sources, black dots the {\it reference} sample. The numbers in each
     panel show the median $R_{h,test}/R_{h,ref}$ ratio, with the
     $rms_{MAD}$ and the number of objects used in parenthesis, labels are
     colour coded.}
   \label{ishape_tests}
   \end{figure}

In order to determine the best parameters for \textsf{Ishape} we
performed a reference run and various tests changing the input
parameters.  Table \ref{tab_ishape} gives the main parameters for the
reference run (\textsf{g1} label). The other tests are obtained as
follows: test\#1 we adopted the DAOPHOT Penny1 PSF instead of the
Moffat25 (label \textsf{g2}); for test\#2 and \#3 (labels \textsf{g3}
and \textsf{g4}) we used \textsf{Ishape} fitting radius 9 pixels and
15 pixels, respectively; test\#4: the maximum FWHM is set to 40 pixels
(label \textsf{g5}); test\#5: does not fit an elliptical model,
circular symmetry is used instead (label \textsf{g6}). In all cases,
except for the test \#6, the FWHM is transformed to circularized
effective radius $R_h=1.48\cdot FWHM_{KING30}\cdot0.5\cdot
(1+w_y/w_x)$, where $FWHM_{KING30}$ and $w_y/w_x$ are the full width,
and the axis ratio fitted by \textsf{Ishape}\footnote{This equation,
  suggested in the \textsf{Ishape} handbook, provides results nearly
  identical to the one $R_h=1.48\cdot FWHM_{KING30}\cdot
  \sqrt{w_y/w_x}$ used by other authors \citep[e.g.][]{blake08}.}. For
the test \#6 we used $R_h=1.48\cdot FWHM_{KING30}$.  Figure
\ref{ishape_tests} shows the results of the \textsf{Ishape} tests. The
data in the figure show that, in general, there can be even a factor
10 difference between $R_h$ estimates with different \textsf{Ishape}
input parameters. Nevertheless, for the {\it reference} sample (see
Section \ref{sec_sizes}), the effect of changing fitting parameters
implies a median change on $R_h$ of $\lsim10\%$. We also inspected the
correlation between the radius and magnitude of the sources, and did
not find any.

\begin{figure}
   \centering
  \resizebox{\hsize}{!}{\includegraphics{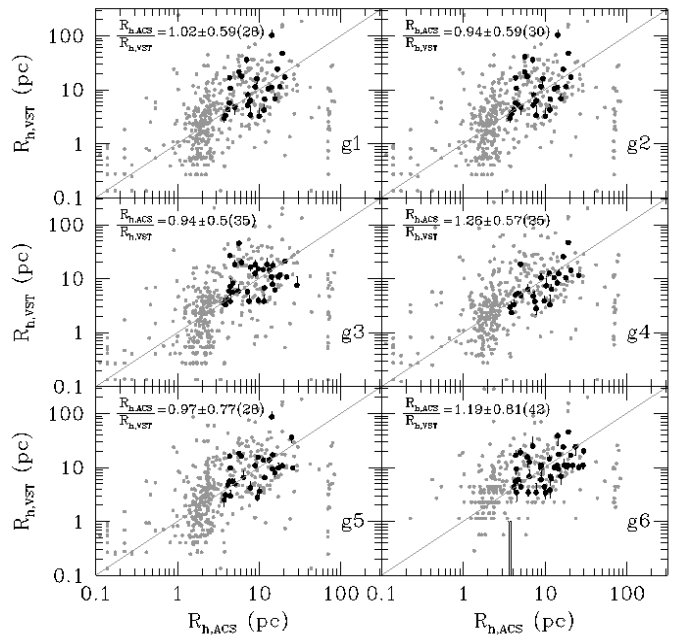}} 
   \caption{As in left panel of Figure \ref{ishape_ref}, but for the different
     \textsf{Ishape} tests (as
     labeled).}
   \label{ishape_acsvst_stat}
   \end{figure}

Figure \ref{ishape_acsvst_stat} shows the \sss~ to ACS comparison for
the various tests. The data in the figure show that the results with
the reference \textsf{g1} run are broadly consistent with the other
tests. We note that, in choosing the best parameters for \textsf{Ishape}, we
also took into account the number of sources successfully analyzed by
the tool. For the \textsf{g1} test the input catalog contained
$\sim47000$ sources, and the spatial parameters were obtained for
$\sim30000$. Such number can decrease significantly for other choices
of the input parameters -- most notably in the test \textsf{g4}.

%____ISHAPE TESTS__________________________________________ 

\section{On some UCDs in \citet{jennings14}}
\label{app_ucd}

\begin{figure}
   \centering
   \resizebox{\hsize}{!}{\includegraphics{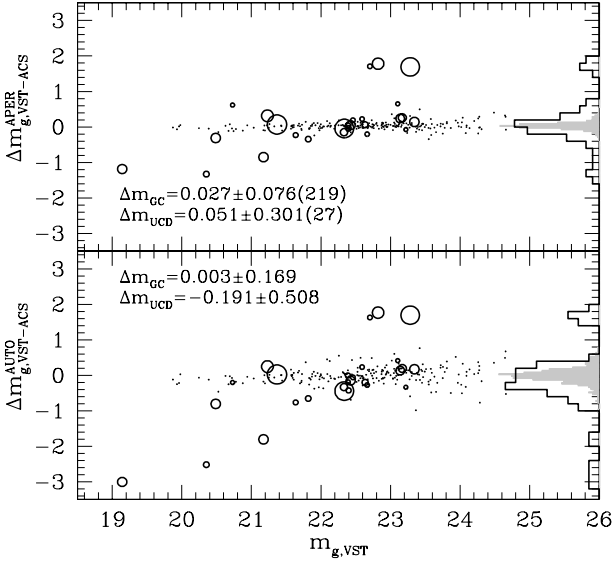}}
   \caption{Upper panel: comparison of ACS and \sss~ aperture corrected
     magnitudes for the GCs and UCDs. Dots and shaded histogram refer
     to GCs, UCDs are shown with empty circles (with symbol size
     scaled to $R_h$) and thick solid line histogram. The number of
     matched sources is reported together with the median and $rms$ of
     the \sss~ to ACS $m_g$ difference. Lower panel: as upper but the
     SExtractor AUTO magnitude is used for both GCs and UCDs.}
   \label{ucds}
   \end{figure}

As discussed in Section \ref{sec_sizes}, and shown in Figure
\ref{acsvstmag}, we found a good match with ACS photometry. Figure
\ref{ucds} shows that the photometric matching is not as good for some
of the UCD candidates in \citeauthor{jennings14} catalog (empty
circles). The mismatch cannot be simply explained by the different
aperture correction, since, as described in \citet[][Section
  2.4.1]{jennings14}, the largest aperture correction is 0.94 mag, and
we find differences up to $\sim2$ mag.

\begin{figure} 
   \centering
 \resizebox{\hsize}{!}{\includegraphics{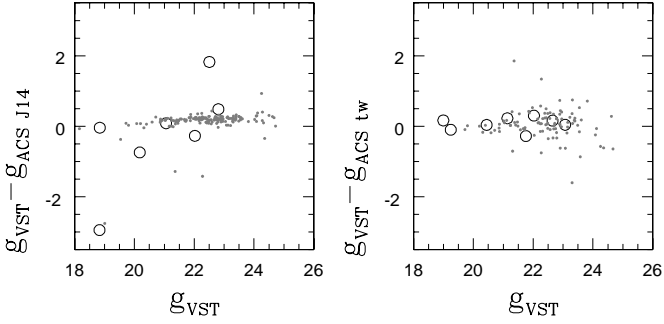}}
  \caption{A comparison of \sss~ $g$-band photometry ($g_{VST}$
    label), with our photometry of ACS data ($g_{ACS~tw}$), and with
    \citet{jennings14} measurements ($g_{ACS~J14}$). UCDs are shown
    with empty circles, GCs with dots. Left panel shows the mismatch
    for UCDs between \sss~ and ACS measurements from
    \citet{jennings14}. Right panel: as left, but our measures for the
    ACS pointings are used. }
   \label{acsvstmag2}
   \end{figure}

To understand what the issue is, we downloaded one of the ACS
pointings of NGC\,3115 (choosing the one maximizing the number of UCDs
over the frame), and independently derived the photometry of SSSs
candidates using the same methods and tools described in Section
\S\ref{sec_photo}. Figure \ref{acsvstmag2} shows the comparison \sss~ magnitudes ($g_{VST}$), with our
photometry from ACS images ($g_{ACS~tw}$), and with 
\citet{jennings14} ($g_{ACS~J14}$). From the latter
 test, we find that, while the agreement for GC photometry is
still acceptable, the large difference between ACS and \sss~ UCD data
disappears (right panel in the figure). The large scatter for the GCs
is mainly due to the use of only one of the ACS pointings available.

\begin{figure} 
   \centering
 \resizebox{\hsize}{!}{\includegraphics{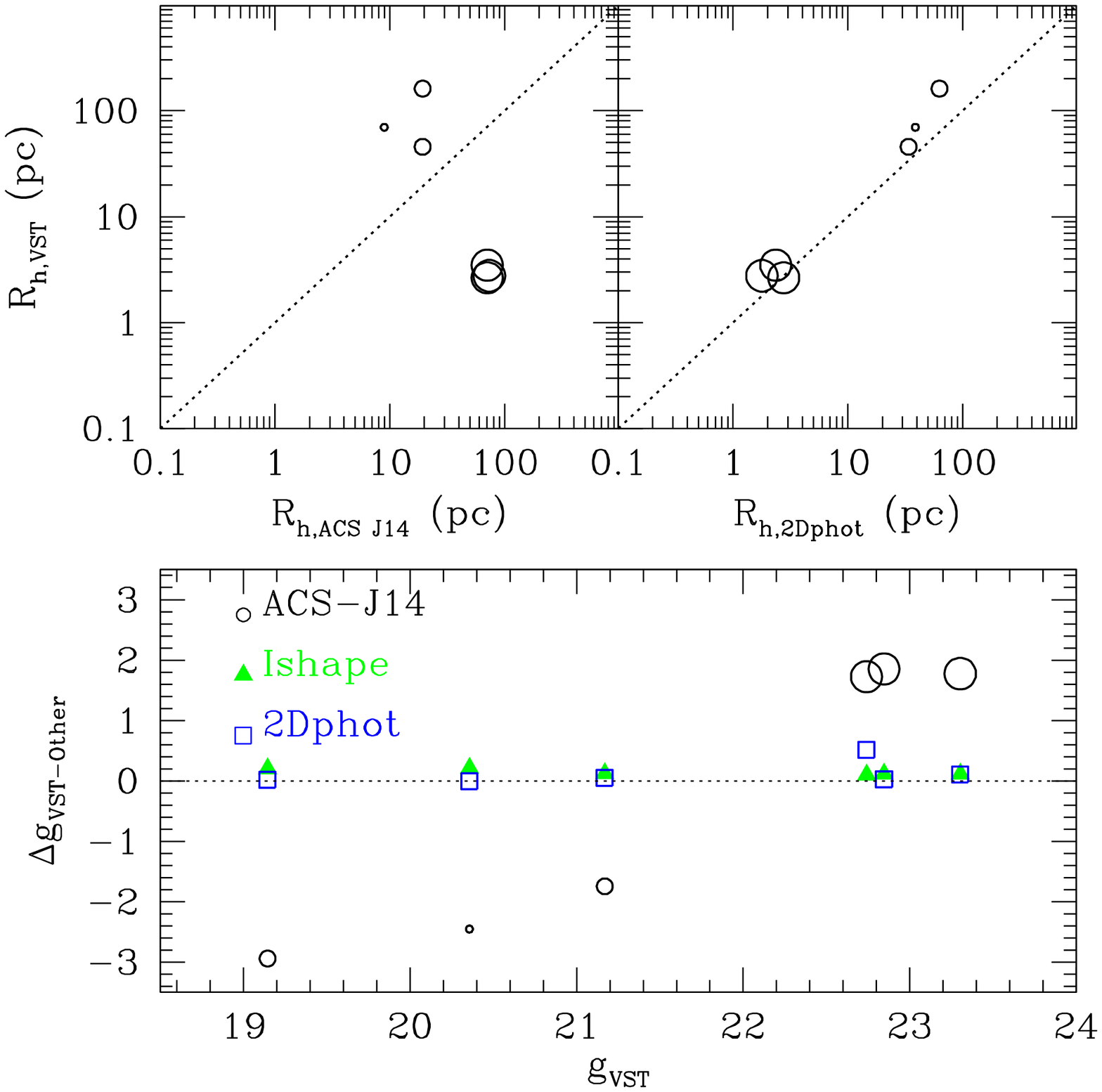}} 
  \caption{Upper panels: comparisons of effective radii for
    UCD candidates obtained by \citet{jennings14} and \sss~
    results. Upper left panel: comparison between \sss~ and ACS sizes from \citet{jennings14}. Upper right
    panel: comparison between the $R_h$ from \sss,
    and measurements with 2Dphot. Lower panel: photometric comparison
    for the same UCDs in upper panels. \sss~ photometry is taken as
    {\it reference}. The difference with respect to ACS data from
    \citet{jennings14}, magnitudes derived with \textsf{Ishape}, and magnitudes
    from 2Dphot are shown with black circles, green triangles, and
    blue empty squares respectively. In all panels symbols size are scaled to the $R_h$ from
    \citet{jennings14}.}
   \label{photo_ucd}
   \end{figure}

As a further check, for the UCDs in common with the ACS pointing
analyzed, we also compared magnitudes and effective radii for \sss,
using a different photometric tool, 2Dphot \citep[][test kindly
  carried out by F. La Barbera]{labarbera08}, and found a good
matching between the results of our standard procedures and the ones
from 2Dphot (Figure \ref{photo_ucd}, upper right and lower panels).

A visual inspection of some UCD candidates from \citet{jennings14}
reveals possible problems with the identification of sources. The
cases shown, infact, reveal that UCD10 (the object with the largest
difference in the Figures \ref{ucds}-\ref{acsvstmag2}) and UCD20 from
the \citeauthor{jennings14} list, are actually a spiral galaxy and an
object immersed in system with clear merging features (tidal
streams?).

%____UCDs comparison with ZJ__________________________________________ 
%

 %___________myACS___________________________ 

 %______2dphot______________________________ 
\clearpage
 %_____BAD UCDs ZJ_________________________________ 
  \begin{figure*} 
   \centering
   \includegraphics[width=12cm]{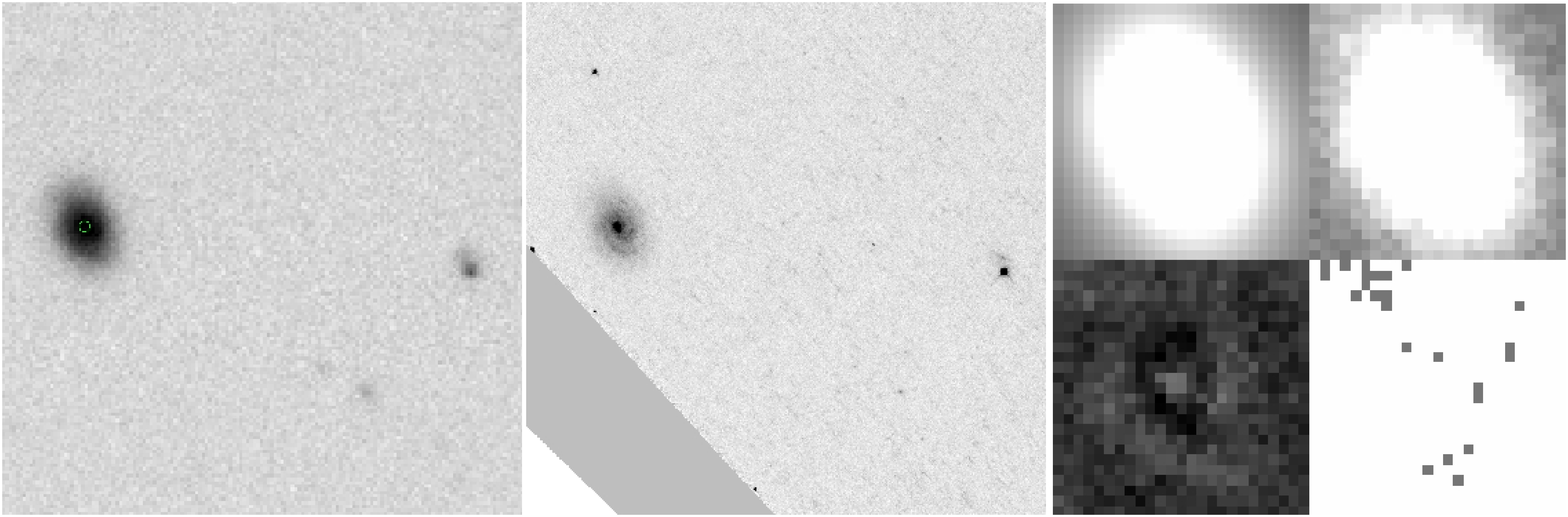} 
   \includegraphics[width=12cm]{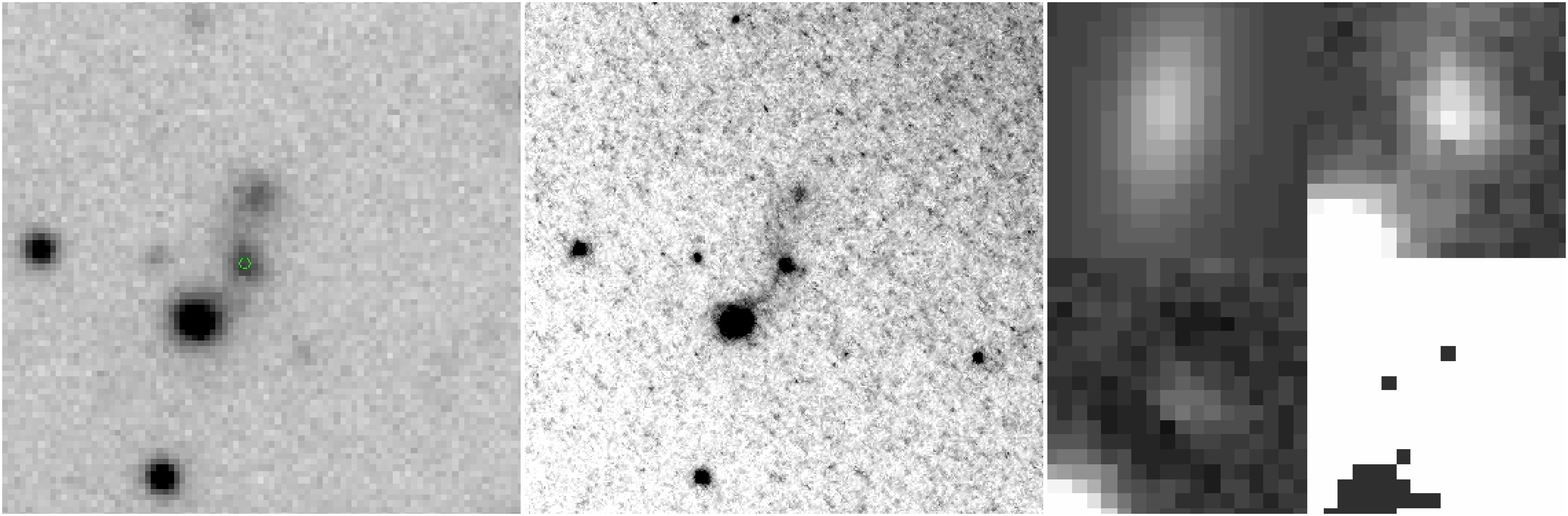} 
  \caption{Two examples of UCD in common with \citet{jennings14}
    showing large photometric scatter with respect to
    \sss. Left/middle panel: VST/ACS $g$ image. Right panel:
    \textsf{Ishape} residual cutout. The UCD candidate is highlighted
    with green circle in the \sss~ panels. Upper/lower panels refer to
    UCD10/UCD20 in the catalog (15$\arcsec$/10$\arcsec$ zoom box,
    respectively).}
   \label{ucd_bad}
   \end{figure*}

\end{appendix}

\end{document}